\documentclass[12pt]{article}
\usepackage[utf8]{inputenc}
\usepackage{amsmath,amsfonts,amssymb,amsthm}
\usepackage[mathcal]{eucal}
\usepackage[longnamesfirst,sort&compress]{natbib}
\usepackage[margin=1in]{geometry}
\usepackage{indentfirst}
\usepackage{setspace,graphicx,pdflscape,datetime,subfigure,verbatim,float,tabularx,booktabs}
\usepackage{hyperref} \hypersetup{pdfpagemode=FullScreen,colorlinks=true,citecolor=blue,urlcolor=blue}
\usepackage[]{caption}  

\newcommand{\comments}[1]{}
\newcommand{\rp}[1]{\footnote{RP-#1}}
\newcolumntype{F}{>{\raggedright\arraybackslash}X}
\newcolumntype{G}{>{\raggedleft\arraybackslash}X}
\newcolumntype{N}{>{\centering\arraybackslash}X}
\newcommand\RPtxtsize[1]{\normalsize{#1}}
\newcommand{\RPprep}[5]{%
  \def\ArgI{#1}%									Caption
  \def\ArgII{#2}%									scale  (0=normal,1=small)
  \def\ArgIII{#3}%								    lscape (0=normal,1=lscape)
  \def\ArgIV{#4}%									label
  \def\ArgV{#5}%									notes
  \ifnum \ArgII=1
    \renewcommand\RPtxtsize[1]{\footnotesize{##1}}
  \else
    \renewcommand\RPtxtsize[1]{\normalsize{##1}}
  \fi
}

\newcommand\RPtab[1]{%
		\ifnum \ArgIII=1
			\begin{landscape}
		\fi
		\begin{table}[!hbp]
		\caption{\ArgI}
		\ArgV
		\vfill
		\centering
		\RPtxtsize{
		#1
		}
		\label{tab:\ArgIV}
	\end{table}
	\ifnum \ArgIII=1
		\end{landscape}
	\fi
}	

\newcommand{\RPfig}[1]{
	\ifnum \ArgIII=1
		\begin{landscape}
	\fi
	\begin{figure}[!hbp]
		\centering
		\RPtxtsize{
		#1
		}
		\caption{\ArgI. %
		    \ArgV
		}
		\vfill
		\medskip
		\label{fig:\ArgIV}
	\end{figure}
	\ifnum \ArgIII=1
		\end{landscape}
	\fi
}

\begin{document}

\title{The Difference-of-Log-Normals Distribution:\protect\\Properties, Estimation, and Growth}

\author{
	Robert Parham\footnote{University of Virginia (robertp@virginia.edu). Code for working with DLN distributions is available from the author.}
}
\date{\today}
\renewcommand{\thefootnote}{\fnsymbol{footnote}}
\maketitle

\vspace{-.2in}
\begin{abstract}
\noindent This paper describes the Difference-of-Log-Normals (DLN) distribution. A companion paper \cite{Parham2022} makes the case that the DLN is a fundamental distribution in nature, and shows how a simple application of the CLT gives rise to the DLN in many disparate phenomena. Here, I characterize its PDF, CDF, moments, and parameter estimators; generalize it to N-dimensions using spherical distribution theory; describe methods to deal with its signature ``double-exponential'' nature; and use it to generalize growth measurement to possibly-negative variates distributing DLN. I also conduct Monte-Carlo experiments to establish some properties of the estimators and measures described.
\end{abstract}

\medskip
\thispagestyle{empty}
\setcounter{footnote}{0}
\renewcommand{\thefootnote}{\arabic{footnote}}
\setcounter{page}{1}
\doublespacing 
%\onehalfspacing 

%%%%%%%%%%%%%%%%%%%%%%%%%%%%%%%%%%%%%%%%%%%%%%%%%%%%%%%%%%%%%%%%%%%%%%%%%%%%%%%%%%%%%%%%%%%%%
% Body
\clearpage

\section{Introduction}

The difference-of-log-Normals distribution, henceforth DLN, is the distribution arising when one subtracts a log-Normal random variable (RV) from another. To define the DLN, consider an RV $W$ such that 
\begin{equation} \label{eq:DLN}
W = Y_{p} - Y_{n} = \text{exp}(X_{p}) - \text{exp}(X_{n}) \ \ \text{with} \ \ \pmb{X} = (X_{p},X_{n})^{T} \sim \mathcal{N}(\pmb{\mu},\pmb{\Sigma})
\end{equation}
in which $\pmb{X}$ is a bi-variate Normal with
\begin{equation} \label{eq:BVN}
\pmb{\mu} = \begin{bmatrix} \mu_p \\ \mu_n \end{bmatrix} \ \ \ 
\pmb{\Sigma} = \begin{bmatrix} \sigma_p^2 & \sigma_p\cdot\sigma_n\cdot\rho_{pn} \\ \sigma_p\cdot\sigma_n\cdot\rho_{pn} & \sigma_n^2 \end{bmatrix}
\end{equation}
We say $W$ follows the five-parameter DLN distribution, $W \sim \text{DLN}(\mu_p,\sigma_p,\mu_n,\sigma_n,\rho_{pn})$.

The companion paper \cite{Parham2022} makes the case that the DLN is a \emph{fundamental distribution in nature}, in the sense that it arises naturally in a plethora of disparate natural phenomena, similar to the Normal and log-Normal distributions. It shows that firm income, return, and growth are all well-described by the DLN, it further shows that city population growth, per-county GDP growth, and the per-industry per-Metro GDP growth all show remarkable fit to the DLN. \cite{Parham2022} describes how the emergence of the DLN is a direct result of an application of the Central Limit Theorems and ``Gibrat's Law'' when applied to various economic phenomena. As the DLN is almost completely unexplored,\footnote{At the time of writing, I was able to find only two statistical works considering it, \cite{Lo2012} and \cite{GulisashviliTankov2016}. Both papers concentrate on the sum of log-Normals but show their results hold for the difference of log-Normals as well, under some conditions.} this paper aims to fill the gap.

The next section fully characterizes the DLN distribution, deriving its PDF, CDF, central moments, and estimators for the distribution parameters given data. It also introduces an extension of the DLN to the multi-variate N-dimensional case using elliptical distribution theory. A full suite of computer code is provided for future use.

Next, Section~\ref{sec:Methods} discusses the difficulty of working with the raw DLN distribution, stemming from its characteristic ``double-exponential'' heavy tails. To alleviate this difficulty, I discuss the close link between the DLN and the Hyperbolic Sine (\emph{sinh}) function and its inverse (\emph{asinh}) and present the ADLN distribution - the DLN under an asinh transform. The section then considers the problem of measuring growth in DLN-distributed RVs. To that end, it generalizes the concept of growth, currently defined only for strictly positive RVs, to DLN RVs that are sometimes negative. I show that the appropriate growth concept for an RV (e.g. percentage, difference in logs, or DLN-growth) intimately depends on the RV's statistical distribution.

Section~\ref{sec:MC} explores the properties of the estimators presented via extensive Monte-Carlo experiments. It: (i) reports the empirical bias and variance of the moment estimators and the MLE parameter estimators; (ii) establishes critical values for the Kolmogorov-Smirnov and Anderson-Darling distributional tests for DLN RVs; and (iii) presents the relation between the measures of growth developed in Section~\ref{sec:Methods}. \comments{Finally, it discusses using the DLN as an approximating distribution, and presents evidence that the DLN is an excellent approximating distribution for several distributions, including the distributions arising from the sum of two DLN RVs, the multiplication of DLN by Normal RVs, and the multiplication of Normal by log-Normal RVs.}

\section{Definitions and properties}
\label{sec:Def}

Prior to proceeding, and to fix ideas, Figure~\ref{fig:DLNexam} presents several instances of the DLN distribution. Panel (a) presents and contrasts the standard Normal, standard DLN, and standard log-Normal. The uncorrelated standard DLN is defined as DLN(0,1,0,1,0), i.e. the difference between two exponentiated uncorrelated standard Normal RVs. Panel (b) shows the role of the correlation coefficient $\rho_{pn}$ in the standard DLN, controlling tail-weight vs. peakedness. Panel (c) repeats the analysis of Panel (b) for a different parametrization common in practical applications, exhibiting the problem of dealing with heavy tails. Panel (d) presents the data of panel (c) in asinh space, showing how asinh resolves the problem of graphing heavy tails and why the ADLN distribution is useful in practice.

% DLN Examples
\RPprep{DLN Examples}{0}{0}{DLNexam}{%
    This figure presents examples of the DLN distribution. Panel (a) graphs the PDFs of the standard Normal, log-Normal, and DLN. Panel (b) graphs the PDFs of standard DLN with different correlation coefficients $\rho_{pn}$. Panel (c) presents the PDFs of a DLN with parameters $(3,2,2,2)$, common in practice, and varying correlation coefficients $\rho_{pn}$. Panel (c) presents the PDF for the range $\pm 10$, which is a significant truncation due to the long tails of this DLN. Panel (d) presents the same PDFs as Panel (c), but the x-axis is asinh-transformed, such that it spans the range sinh(-10) $\approx$ -11,000 to sinh(10) $\approx$ 11,000.
}
\RPfig{%
	\begin{tabular}{cc} 
		\subfigure[standard DLN, N, LN]{\includegraphics[width=3in]{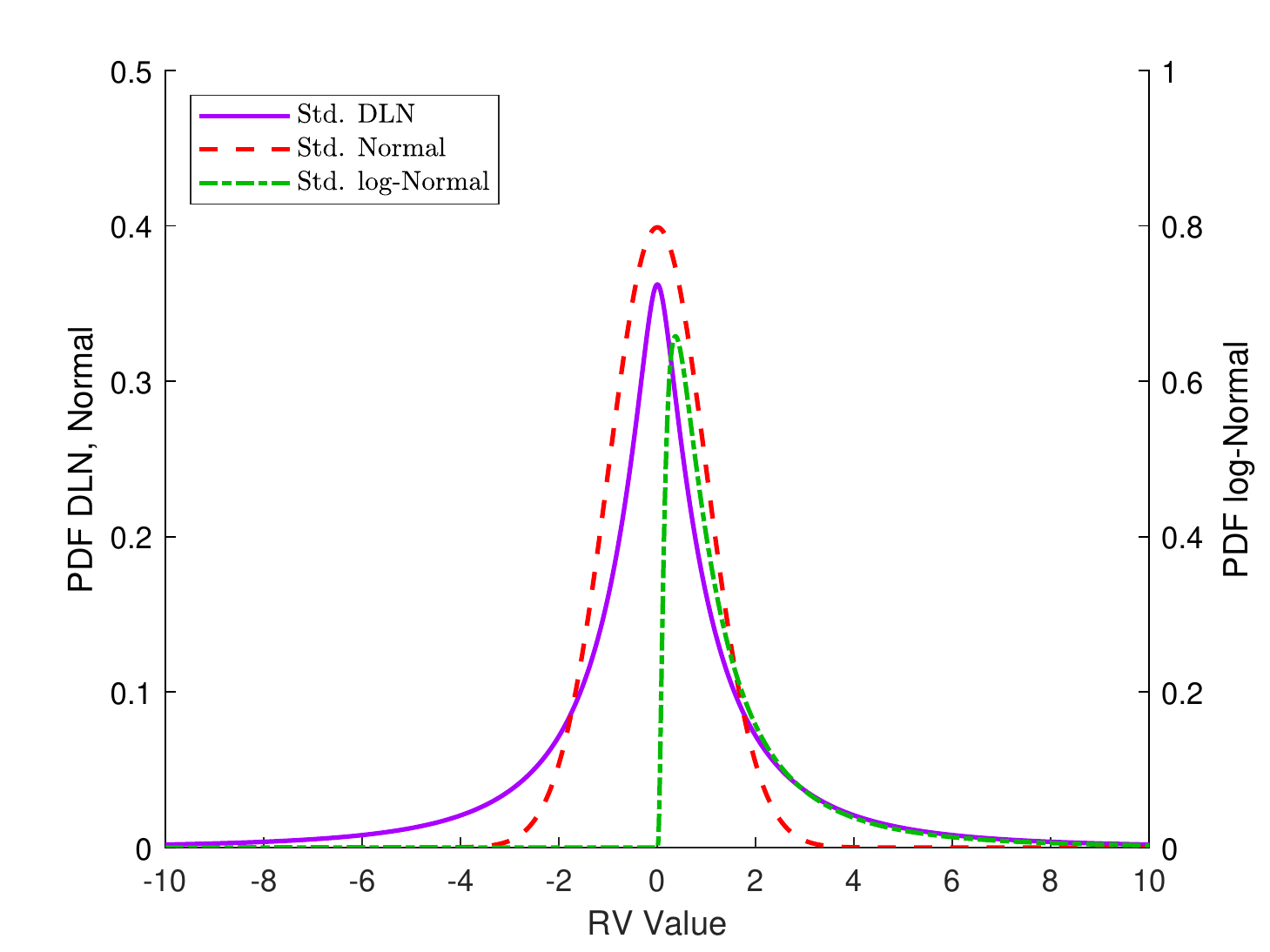}} & 
		\subfigure[Std. DLN w/ corrs]{\includegraphics[width=3in]{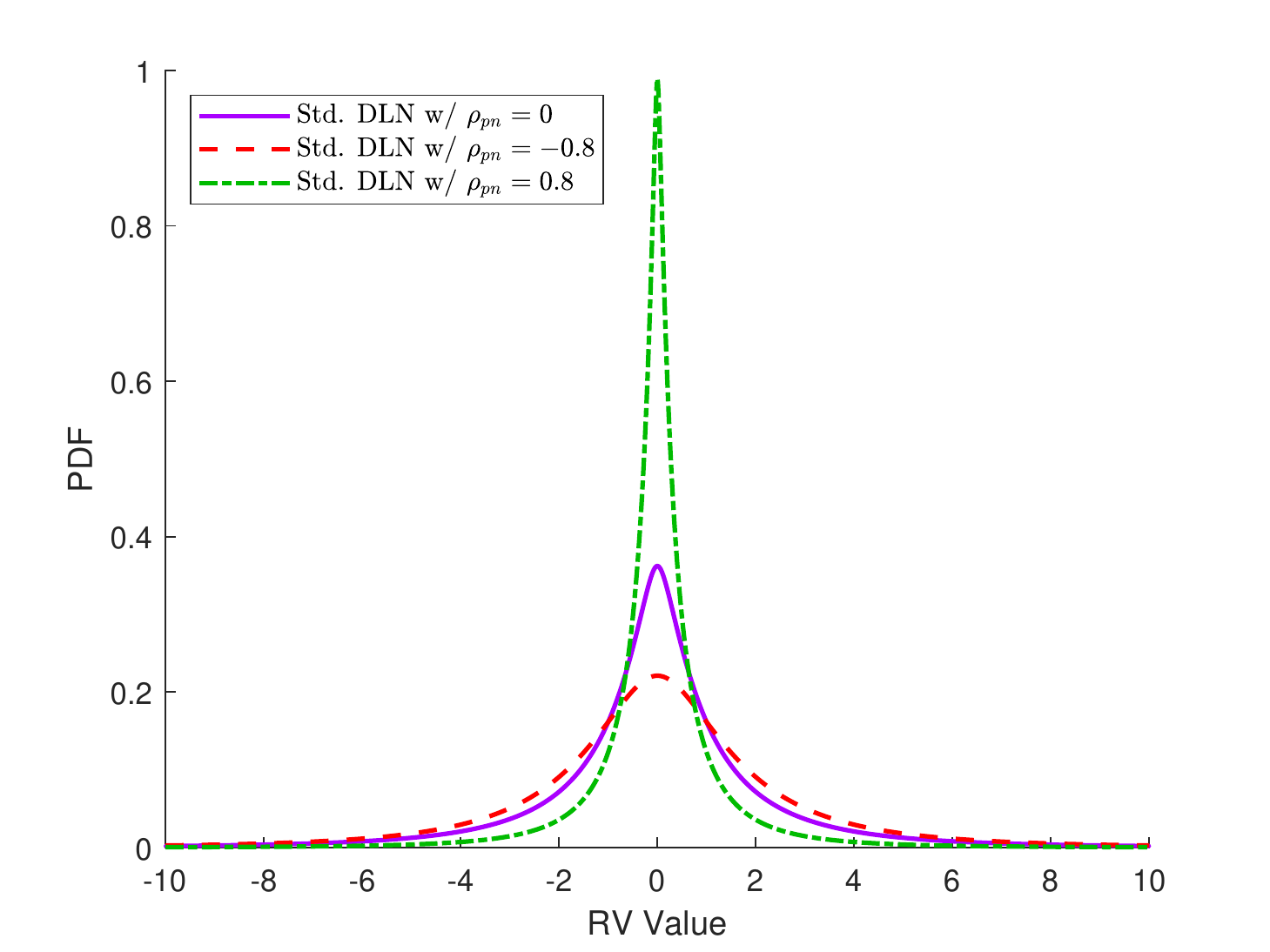}} \\ \\
		\subfigure[DLN w/ corrs]{\includegraphics[width=3in]{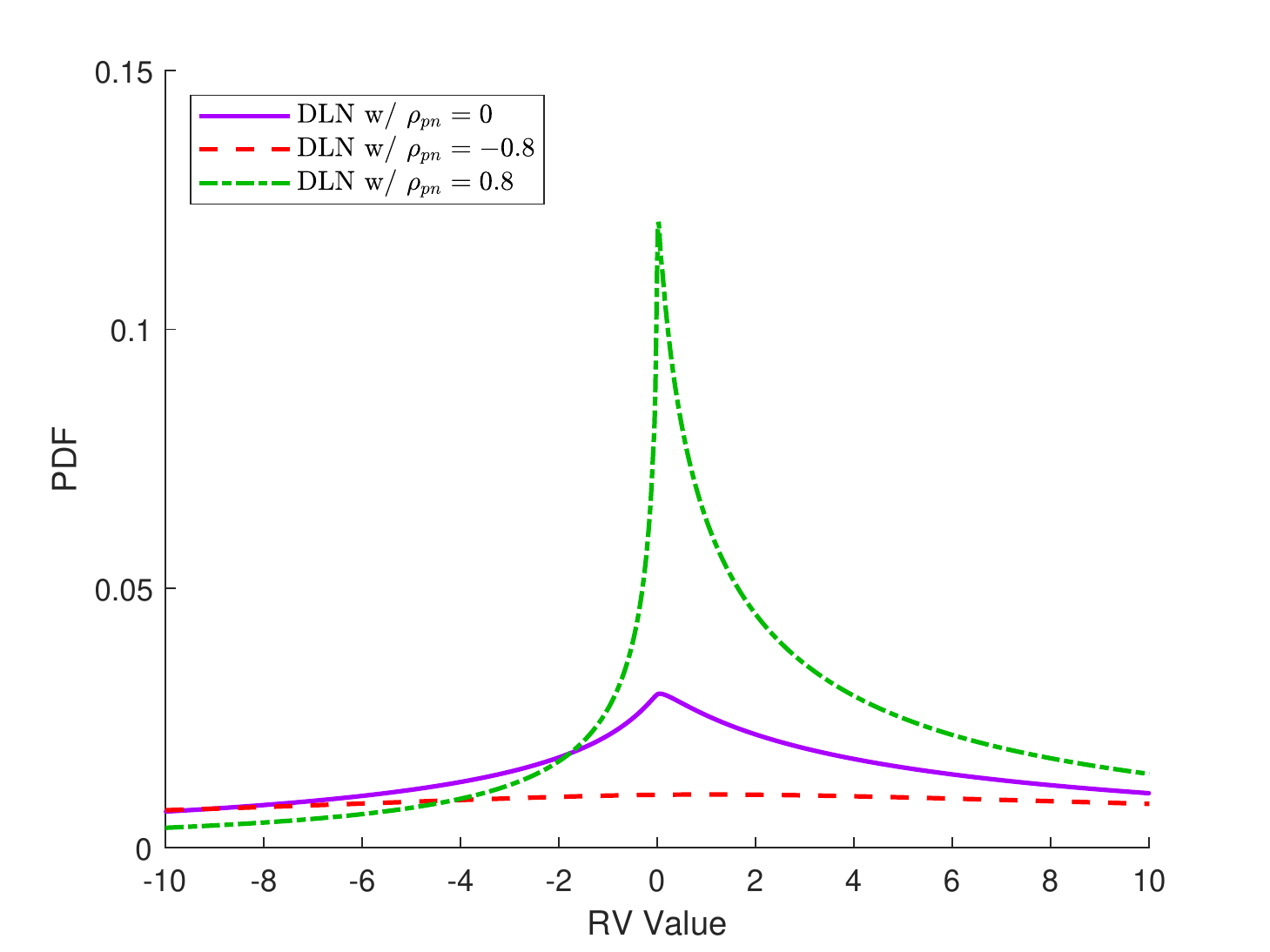}} &
		\subfigure[ADLN w/ corrs]{\includegraphics[width=3in]{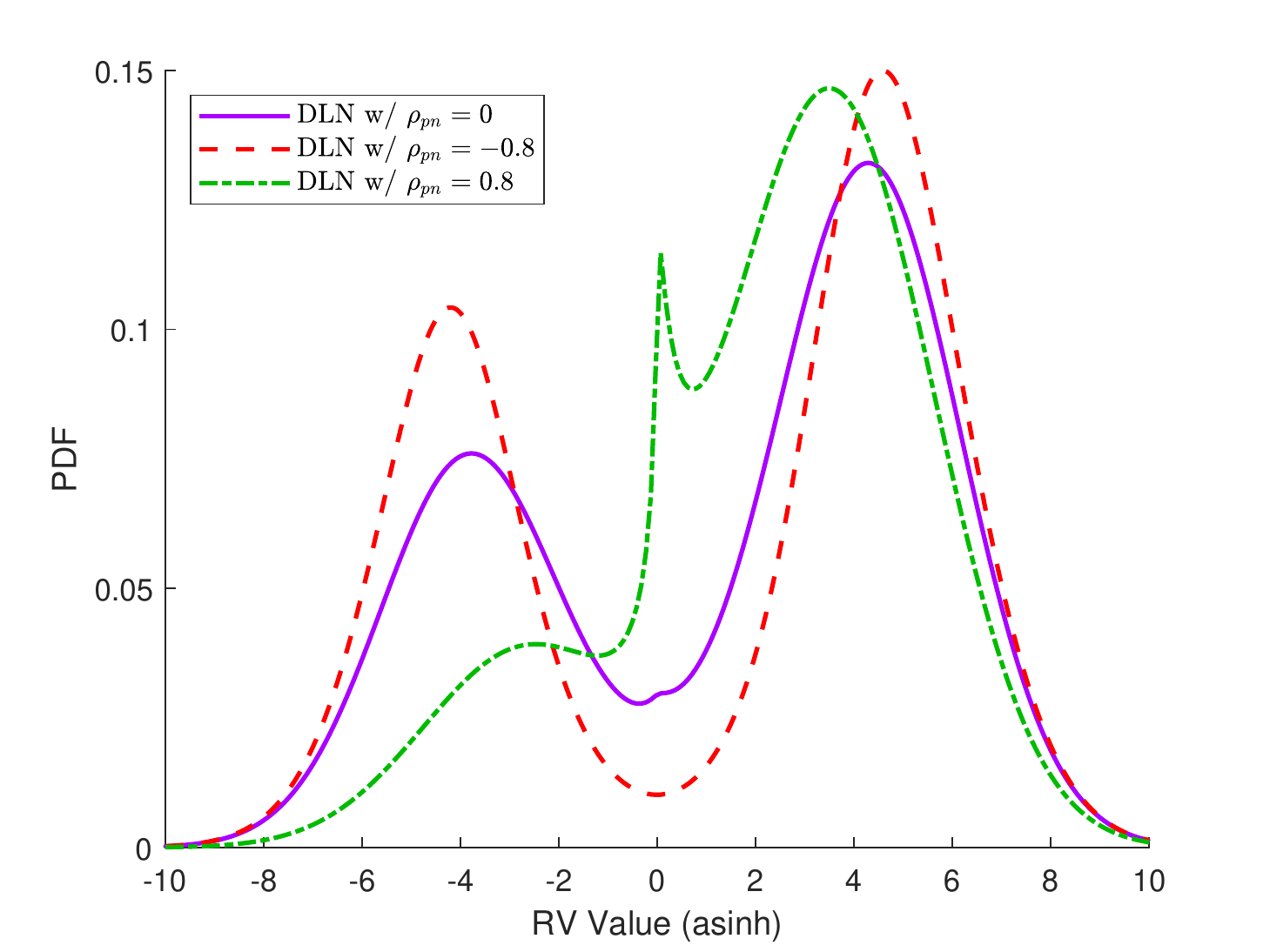}} \\ \\
	\end{tabular}
}

\subsection{PDF and CDF}

The PDF for the bi-variate Normal (BVN) RV $\pmb{X}$ is well-known to be
\begin{equation} \label{eq:PDFBVN}
f_{BVN}(\pmb{x}) = \frac{\lvert\pmb{\Sigma}\rvert^{-\frac{1}{2}}}{2\pi}\cdot \text{exp}\left(-\frac{1}{2} (\pmb{x}-\pmb{\mu})^{T} \pmb{\Sigma}^{-1} (\pmb{x}-\pmb{\mu})\right) = \frac{\lvert\pmb{\Sigma}\rvert^{-\frac{1}{2}}}{2\pi}\cdot \text{exp}\left(-\frac{1}{2} \lvert\lvert\pmb{x}-\pmb{\mu}\rvert\rvert_{\pmb{\Sigma}}\right)
\end{equation} 
with $\lvert\pmb{\Sigma}\rvert$ the determinant of $\pmb{\Sigma}$ and $\lvert\lvert\pmb{x}\rvert\rvert_{\pmb{\Sigma}}$ the Euclidean norm of $\pmb{x}$ under the Mahalanobis distance induced by $\pmb{\Sigma}$.

The PDF for the bi-variate log-Normal (BVLN) RV $\pmb{Y} = (Y_{p},Y_{n})^{T}$ can be obtained by using the multivariate change of variables theorem. If $\pmb{Y}=g(\pmb{X})$ then
\begin{equation}
f_{Y}(\pmb{y}) = f_{X}(g^{-1}(\pmb{y})) \cdot \lvert\lvert J_{g^{-1}}(\pmb{y})\rvert\rvert
\end{equation}
with $J_{g^{-1}}$ the Jacobian matrix of $g^{-1}(\cdot)$ and $\lvert\lvert J_{g^{-1}}\rvert\rvert$ the absolute value of its determinant. Applying the theorem for $\pmb{Y} = g(\pmb{X}) = (\text{exp}(X_p),\text{exp}(X_n))^{T}$ we have $g^{-1}(\pmb{y}) = (log(y_p),log(y_n))^{T}$ and $\lvert\lvert J_{g^{-1}}(\pmb{y})\rvert\rvert = (y_p\cdot y_n)^{-1}$. The PDF of a BVLN RV is then
\begin{equation} \label{eq:PDFBVLN}
f_{BVLN}(\pmb{y}) = \frac{\lvert\pmb{\Sigma}\rvert^{-\frac{1}{2}}}{2\pi y_p y_n} \text{exp}\left(-\frac{1}{2}\lvert\lvert\log(\pmb{y})-\pmb{\mu}\rvert\rvert_{\pmb{\Sigma}}\right)
\end{equation}

We can now define the cumulative distribution function (CDF) of the DLN distribution using the definition of the CDF of the difference of two RV
\begin{equation} \label{eq:CDFDLN1}
\begin{split}
F_{DLN}(w) & = P[W\leq w] = P[y_{p} - y_{n} \leq w] = P[y_{p} \leq y_{n} + w] \\
           & =  \int_{-\infty}^{\infty}\int_{-\infty}^{y_{n}+w}f_{BVLN}(y_{p},y_{n})dy_{p}dy_{n}
\end{split}
\end{equation}
which can be differentiated w.r.t $w$ to yield the PDF
\begin{equation} \label{eq:PDFDLN1}
f_{DLN}(w) = \int_{-\infty}^{\infty}f_{BVLN}(y+w,y)dy  =  \int_{-\infty}^{\infty}f_{BVLN}(y,y-w)dy
\end{equation}
but because $f_{BVLN}(\pmb{y})$ is non-zero only for $\pmb{y}>0$, we limit the integration range
\begin{equation} \label{eq:PDFDLN}
f_{DLN}(w) = \int_{\text{max}(0,w)}^{\infty}f_{BVLN}(y,y-w)dy
\end{equation}
which yields the PDF of the DLN distribution.

It is well-known, however, that the integral in equation~\ref{eq:PDFDLN} does not have a closed-form solution. The accompanying code suite evaluates it numerically, and also numerically evaluates the CDF using its definition
\begin{equation} \label{eq:CDFDLN}
F_{DLN}(w) = \int_{-\infty}^{w}f_{DLN}(y)dy
\end{equation}

\begin{sloppypar}
For the simpler case with difference of uncorrelated log-Normals, i.e. $\rho_{pn}=0$, we can derive the PDF of the DLN via a characteristic function (CF) approach as well. In this case, we can write the CF of the DLN as ${\varphi_{DLN}(t)=\varphi_{LN}(t)\cdot\varphi_{LN}(-t)}$ with $\varphi_{LN}(t)$ the CF of the log-Normal. Next, we can apply a Fourier transform to obtain the PDF,
\begin{equation} \label{eq:PDFDLNCF}
f_{DLN}(w) = \frac{1}{2\pi}\int_{-\infty}^{\infty}e^{-i\cdot t\cdot w} \cdot \varphi_{DLN}(t)dt
\end{equation}
Unfortunately, the log-Normal does not admit an analytical CF, and using Equation~\ref{eq:PDFDLNCF} requires a numerical approximation for $\varphi_{LN}(t)$ as well. \cite{Gubner2006} provides a fast and accurate approximation method for the CF of the log-Normal which I use in the calculation of $f_{DLN}(w)$ when using this method.
\end{sloppypar}

\subsection{Moments}
\label{sec:Moms}

\subsubsection{MGF}

The moment generating function (MGF) of the DLN can be written as
\begin{equation} \label{eq:MGFDLN}
M_{W}(t) = \mathbb{E}\left[e^{tW}\right] = \int_{-\infty}^{\infty}\int_{-\infty}^{\infty}e^{tw}f_{BVLN}(y+w,y)dydw
\end{equation}
but this formulation has limited usability due to the lack of closed-form solution for the integrals. Instead, it is useful to characterize the moments directly, as we can obtain them in closed-form.

\subsubsection{Mean and variance}

Using the definitions of $\pmb{\mu}$ and $\pmb{\Sigma}$ in \ref{eq:BVN}, define the mean and covariance of the BVLN RV, $\pmb{\hat{\mu}}$ and $\pmb{\hat{\Sigma}}$ (element-wise) as
\begin{equation} \label{eq:BVLN}
\begin{split}
\pmb{\hat{\mu}}_{(i)} & = \text{exp}\left(\pmb{\mu}_{(i)} + \frac{1}{2}\pmb{\Sigma}_{(i,i)}\right) \\
\pmb{\hat{\Sigma}}_{(i,j)} & = \text{exp}\left(\pmb{\mu}_{(i)} + \pmb{\mu}_{(j)} + \frac{1}{2}\left(\pmb{\Sigma}_{(i,i)} + \pmb{\Sigma}_{(j,j)}\right)\right)\cdot \left( \text{exp}\left(\pmb{\Sigma}_{(i,j)}\right)-1\right) \\
\end{split}
\end{equation}
Note that if $\pmb{\Sigma}$ is diagonal (i.e., $X_{p}$ and $X_{n}$ are uncorrelated) then $\pmb{\hat{\Sigma}}$ will be diagonal as well. We are however interested in the general form of the DLN distribution. The identities regarding the expectation and variance of a sum of RV yield
\begin{equation} \label{eq:MUDLN}
\mathbb{E}\left[W\right] = \mathbb{E}\left[Y_p\right] - \mathbb{E}\left[Y_n\right] = \pmb{\hat{\mu}}_{(1)} - \pmb{\hat{\mu}}_{(2)} = \text{exp}(\mu_p + \frac{\sigma_p^2}{2}) - \text{exp}(\mu_n + \frac{\sigma_n^2}{2})
\end{equation}
and
\begin{equation} \label{eq:SIGDLN}
\begin{split}
\text{Var}\left[W\right] & =\mathbb{C}\left[Y_p,Y_p\right] + \mathbb{C}\left[Y_n,Y_n\right] -2\cdot\mathbb{C}\left[Y_p,Y_n\right] = \pmb{\hat{\Sigma}}_{(1,1)} + \pmb{\hat{\Sigma}}_{(2,2)} - 2\cdot\pmb{\hat{\Sigma}}_{(1,2)} \\
 & = \text{exp}\left(2\mu_{p}+\sigma_p^2\right)\cdot\left(exp\left(\sigma_p^2\right) - 1\right) 
   + \text{exp}\left(2\mu_{n}+\sigma_n^2\right)\cdot\left(exp\left(\sigma_n^2\right) - 1\right) \\
 & - 2\text{exp}\left(\mu_{p}+\mu_{n}+\frac{1}{2}(\sigma_p^2+\sigma_n^2)\right)
 \cdot\left(\text{exp}\left(\sigma_p\sigma_n\rho_{pn}\right) - 1\right)
\end{split}
\end{equation}
with $\mathbb{C}$ the covariance operator of two general RV $U_{1},U_{2}$
\begin{equation} \label{eq:COVAR}
\mathbb{C}\left[U_{1},U_{2}\right] = \mathbb{E}\left[(U_1 - \mu_1)(U_2-\mu_2)\right]
\end{equation}

\subsubsection{Skewness and kurtosis}

Skewness and kurtosis of the DLN can similarly be established using coskewness and cokurtosis (for overview, see e.g. \cite{Miller2013}). Coskewness of three general RV $U_{1},U_{2},U_{3}$ is defined as
\begin{equation} \label{eq:COSKEW}
\mathbb{S}\left[U_{1},U_{2},U_{3}\right] = \frac{\mathbb{E}\left[(U_1 - \mu_1)(U_2-\mu_2)(U_3-\mu_3)\right]}{\sigma_1\sigma_2\sigma_3}
\end{equation}
and cokurtosis of four general RV $U_{1},U_{2},U_{3},U_{4}$ is defined as 
\begin{equation} \label{eq:COKURT}
\mathbb{K}\left[U_{1},U_{2},U_{3},U_{4}\right] = \frac{\mathbb{E}\left[(U_1 - \mu_1)(U_2-\mu_2)(U_3-\mu_3)(U_4-\mu_4)\right]}{\sigma_1\sigma_2\sigma_3\sigma_4}
\end{equation}
with the property that $\mathbb{S}\left[U,U,U\right] = \text{Skew}\left[U\right]$ and $\mathbb{K}\left[U,U,U,U\right] = \text{Kurt}\left[U\right]$. More importantly, it is simple to show that
\begin{equation} \label{eq:SKEWDIFF}
\text{Skew}\left[U-V\right] = \frac{\sigma_U^3\mathbb{S}\left[U,U,U\right] -3\sigma_U^2\sigma_V\mathbb{S}\left[U,U,V\right]+3\sigma_U\sigma_V^2\mathbb{S}\left[U,V,V\right] -\sigma_V^3\mathbb{S}\left[V,V,V\right]}{\sigma_{U-V}^{3}}
\end{equation}
and similarly
\begin{equation} \label{eq:KURTDIFF}
\begin{split}
\text{Kurt}\left[U-V\right] & = \frac{1}{\sigma_{U-V}^{4}} [ \sigma_U^4\mathbb{K}\left[U,U,U,U\right] -4\sigma_U^3\sigma_V\mathbb{K}\left[U,U,U,V\right] \\ & + 6\sigma_U^2\sigma_V^2\mathbb{K}\left[U,U,V,V\right] -4\sigma_U\sigma_V^3\mathbb{K}\left[U,V,V,V\right]+\sigma_V^4\mathbb{K}\left[V,V,V,V\right] ]
\end{split}
\end{equation}
with $\sigma_{U-V} = \text{Var}\left[U-V\right]^{\frac{1}{2}}$ calculated using Equation~\ref{eq:SIGDLN}. Evaluating the operators $\mathbb{S}$ and $\mathbb{K}$ for the case of DLN requires evaluating expressions of the general form $\mathbb{E}\left[Y_{p}^{i}Y_{n}^{j}\right]$, which can be done via the MGF of the BVN distribution
\begin{equation} \label{eq:EUVSimp}
\mathbb{E}\left[Y_{p}^{i}Y_{n}^{j}\right] = \mathbb{E}\left[e^{i X_p}e^{j X_n}\right] = \text{MGF}_{BVN}\left(\big[\begin{smallmatrix} i \\ j \end{smallmatrix}\big] \right) = \mathbb{E}\left[Y_{p}^{i}\right]\mathbb{E}\left[Y_{n}^{j}\right]e^{ij\pmb{\Sigma}_{(1,2)}}
\end{equation}
with $\mathbb{E}\left[Y_{p}^{i}\right]=\text{exp}\left(i\mu_{p} + \frac{1}{2}i^2\sigma_{p}^2\right)$. This concludes the technical details of the derivation. 

The method presented can be extended to higher central moments as well. The accompanying code suite includes functions that implement the equations above and use them to calculate the first five moments of the DLN given the parameters $(\mu_p,\sigma_p,\mu_n,\sigma_n,\rho_{pn})$. Section~\ref{sec:MC} later describes the results of Monte-Carlo experiments testing the empirical variance and bias of the moments as a function of sample size.

\subsection{Estimation}
\label{sec:Estim}

Given data $\pmb{D} \sim \text{DLN}(\pmb{\Theta})$ with $\pmb{\Theta} = (\mu_p,\sigma_p,\mu_n,\sigma_n,\rho_{pn})$, we would like to find an estimate $\pmb{\hat{\Theta}}$ to the parameter vector $\pmb{\Theta}$. Experiments show that given an appropriate initial guess, the MLE estimates of $\pmb{\Theta}$ perform well in practice. The main parameter of difficulty is $\rho_{pn}$. This parameter is akin to the shape parameter in the Stable distribution, which plays a similar role and is similarly difficult to estimate, see e.g. \cite{FamaRoll1971}. It hence requires special care in the estimation.

The estimation code provided minimizes the negative log-likelihood of the data w.r.t the DLN PDF using a multi-start algorithm. The starting values for the first four parameters are fixed for all start points as:
\begin{equation} \label{eq:ESTIM_GUESS}
\begin{bmatrix}
\mu_p \\ \sigma_p \\ \mu_n \\ \sigma_n
\end{bmatrix} = 
\begin{bmatrix}
\text{Median}\left[\text{log}\left(\pmb{D}\right)\right] \ \ \text{for} \ \ \pmb{D}>0 \\
\text{IQR}\left[\text{log}\left(\pmb{D}\right)\right]/1.35 \ \ \text{for} \ \ \pmb{D}>0  \\
\text{Median}\left[\text{log}\left(-\pmb{D}\right)\right] \ \ \text{for} \ \ \pmb{D}<0  \\
\text{IQR}\left[\text{log}\left(-\pmb{D}\right)\right]/1.35 \ \ \text{for} \ \ \pmb{D}<0  \\
\end{bmatrix}
\end{equation}
while the initial guesses for $\rho_{pn}$ are $(-0.8,-0.3,0,0.3,0.8)$. The estimator $\pmb{\hat{\Theta}}$ is then the value which minimizes the negative log-likelihood in the multi-start algorithm. The estimator inherits asymptotic normality, consistency, and efficiency properties from the general M-estimator theory, as the dimension of $\pmb{\hat{\Theta}}$ is fixed, the likelihood is smooth, and is supported on $\mathbb{R}\ \forall \pmb{\hat{\Theta}}$. A better estimation procedure for the parameters of the DLN might be merited, but is left for future work.

\subsection{The elliptical multi-variate DLN}
\label{sec:mvsdln}

Practical applications of the DLN require the ability to work with multi-variate DLN RVs. I hence present an extension of the DLN to the multi-variate case using elliptical distribution theory, with the standard reference being \cite{FangEtAl1990}.

\begin{sloppypar}
The method of elliptical distributions requires a symmetric baseline distribution. We will therefore focus our attention on the symmetric DLN case in which ${\mu_p=\mu_n\equiv\mu}$ and ${\sigma_p = \sigma_n\equiv\sigma}$, yielding the three parameter uni-variate symmetric distribution $\text{SymDLN}(\mu,\sigma,\rho)=\text{DLN}(\mu,\sigma,\mu,\sigma,\rho)$. I begin by defining a standardized N-dimensional elliptical DLN RV using SymDLN and the spherical decomposition of \cite{CambanisEtAl1981}, and later extend it to a location-scale family of distributions.
\end{sloppypar}

Let $\mathbf{U}$ be an N-dimensional RV distributed uniformly on the unit hyper-sphere in $\mathbb{R}^{N}$ and arranged as a column vector. Let $R\geq0$ be a uni-variate RV independent of $\mathbf{U}$ with PDF $f_{R}\left(r\right)$ to be derived momentarily, and let $\mathbf{Z}=R\cdot\mathbf{U}$ be a standardized N-dimensional elliptical DLN RV. A common choice for $\mathbf{U}$ is $\widehat{\mathbf{U}}/\lvert\lvert\widehat{\mathbf{U}}\rvert\rvert_{2}$ with $\widehat{\mathbf{U}} \sim MVN(\mathbf{0}_N,\mathbf{1}_N)$. $\mathbf{U}$ captures a direction in $\mathbb{R}^{N}$, and we have $\sqrt{\mathbf{U}^{T}\cdot\mathbf{U}} =  \lvert\lvert\mathbf{U}\rvert\rvert_{2} \equiv 1$, which implies $\sqrt{\mathbf{Z}^{T}\cdot\mathbf{Z}} =  \lvert\lvert\mathbf{Z}\rvert\rvert_{2} = R$. We further know that the surface area of an N-sphere with radius $R$ is given by
\begin{equation} \label{eq:Surface}
S_{N}\left(R\right) = \frac{2\cdot\pi^{\frac{N}{2}}}{\Gamma\left(\frac{N}{2}\right)}\cdot R^{N-1}
\end{equation}
and can hence write the PDF of $\mathbf{Z}$ as
\begin{equation} \label{eq:fZPDF1}
f_{\mathbf{Z}}\left(\mathbf{z}\right) = \frac{f_{R}\left(\lvert\lvert\mathbf{z}\rvert\rvert_{2}\right)}{S_{N}\left(\lvert\lvert\mathbf{z}\rvert\rvert_{2}\right)} =  \frac{\Gamma\left(\frac{N}{2}\right)\cdot f_{R}\left(\lvert\lvert\mathbf{z}\rvert\rvert_{2}\right)}{2\cdot\pi^{\frac{N}{2}}\cdot\lvert\lvert\mathbf{z}\rvert\rvert_{2}^{N-1}}
\end{equation}

We require $f_{R}\left(r\right)$ and $f_{\mathbf{Z}}\left(\mathbf{z}\right)$ to be valid PDFs, which yields the conditions
\begin{equation} \label{eq:RZcond}
\begin{split}
& f_{R}\left(r\right) \geq 0\ \forall\ r\in\mathbb{R} \\
& f_{\mathbf{Z}}\left(\mathbf{z}\right) \geq 0\ \forall\ \mathbf{z}\in\mathbb{R}^{N} \\
& \int_{-\infty}^{\infty}f_{R}\left(r\right)\ dr = 1 \\
& \int_{-\infty}^{\infty}\cdot\cdot\cdot \int_{-\infty}^{\infty} f_{\mathbf{Z}}\left(\mathbf{z}\right)\  d\mathbf{z}_{(N)}\cdot\cdot\cdot d\mathbf{z}_{(1)} = 1 \\
\end{split}
\end{equation}
to those, we can add the condition that the properly normalized distribution of $f_{R}\left(r\right)$ will be SymDLN,
\begin{equation} \label{eq:fRcond}
f_{R}\left(r\right) = \widetilde{M}_{N}\left(r\right)\cdot f_{DLN}(r)
\end{equation}
with $\widetilde{M}_{N}\left(r\right)$ chosen such that the conditions in Equation~\ref{eq:RZcond} hold. Solving for this set of conditions yields
\begin{equation} \label{eq:fR}
f_{R}\left(r\right) = \frac{r^{N-1}} {\int_{0}^{\infty}\widetilde{r}^{N-1}\cdot f_{DLN}\left(\widetilde{r}\right)\ d\widetilde{r}}\cdot f_{DLN}\left(r\right)
\end{equation}
and
\begin{equation} \label{eq:fZ}
f_{\mathbf{Z}}\left(\mathbf{z}\right) =  \frac{\Gamma\left(\frac{N}{2}\right)}{2\cdot\pi^{\frac{N}{2}}\cdot \int_{0}^{\infty}\widetilde{r}^{N-1}\cdot f_{DLN}\left(\widetilde{r}\right)\ d\widetilde{r}}\cdot f_{DLN}\left(\lvert\lvert\mathbf{z}\rvert\rvert_{2}\right) = M_{N}\cdot f_{DLN}\left(\lvert\lvert\mathbf{z}\rvert\rvert_{2}\right)
\end{equation}
with $M_{N}$ a normalization constant depending only on the dimension N and the parameters of the baseline SymDLN$\left(\mu, \sigma, \rho\right)$ being used. We can further use the definition of the CDF of $\mathbf{Z}$ to write
\begin{equation} \label{eq:FZ}
\begin{split}
F_{\mathbf{Z}}\left(\mathbf{z}\right) & = \int_{-\infty}^{\mathbf{z}_{(1)}}\cdot\cdot\cdot \int_{-\infty}^{\mathbf{z}_{(N)}} f_{\mathbf{Z}}\left(\mathbf{\widehat{z}}\right)\  d\mathbf{\widehat{z}}_{(N)}\cdot\cdot\cdot d\mathbf{\widehat{z}}_{(1)} \\ 
& = \int_{-\infty}^{\mathbf{z}_{(1)}}\cdot\cdot\cdot \int_{-\infty}^{\mathbf{z}_{(N)}} M_{N}\cdot f_{\mathbf{DLN}}\left(\lvert\lvert\mathbf{z}\rvert\rvert_{2}\right)\  d\mathbf{\widehat{z}}_{(N)}\cdot\cdot\cdot d\mathbf{\widehat{z}}_{(1)} \\
\end{split}
\end{equation}
which concludes the characterization of the standardized N-dimensional
elliptical DLN RV.

Extending the standardized N-dimensional DLN to a location-scale family of distributions is now straightforward. Let $\widetilde{\pmb{\mu}}=\left(\mu_1 , \mu_2 , ... , \mu_N\right)^{T}$ be a column vector of locations and let $\widetilde{\pmb{\Sigma}}$ be a positive-semidefinite scaling matrix of rank $N$. Define 
\begin{equation} \label{eq:MVDLN}
\mathbf{W} = \widetilde{\pmb{\mu}} + \widetilde{\pmb{\Sigma}}^{\frac{1}{2}}\cdot\mathbf{Z}
\end{equation}
with $\widetilde{\pmb{\Sigma}}^{\frac{1}{2}}$ denoting the eigendecomposition of $\widetilde{\pmb{\Sigma}}$. The PDF of $\mathbf{W}$ is then given by
\begin{equation} \label{eq:PDFMVDLN}
\begin{split}
f_\mathbf{W}\left(\mathbf{w}\right) & = \lvert\widetilde{\pmb{\Sigma}}\rvert^{-\frac{1}{2}}\cdot f_{\mathbf{Z}}\left(\widetilde{\pmb{\Sigma}}^{-\frac{1}{2}}\cdot\left(\mathbf{w}-\widetilde{\pmb{\mu}}\right)\right) \\
& = \lvert\widetilde{\pmb{\Sigma}}\rvert^{-\frac{1}{2}}\cdot M_{N}\cdot f_{DLN}\left(\sqrt{\left(\mathbf{w}-\widetilde{\pmb{\mu}}\right)^{T}\cdot\widetilde{\pmb{\Sigma}}^{-1}\cdot\left(\mathbf{w}-\widetilde{\pmb{\mu}}\right)}\right) \\
& = \lvert\widetilde{\pmb{\Sigma}}\rvert^{-\frac{1}{2}}\cdot M_{N}\cdot f_{DLN}\left(\lvert\lvert\mathbf{w}-\widetilde{\pmb{\mu}}\rvert\rvert_{\widetilde{\pmb{\Sigma}}}\right) \\
\end{split}
\end{equation}
The CDF of $\mathbf{W}$ can similarly be written as
\begin{equation} \label{eq:CDFMVDLN}
\begin{split}
F_{\mathbf{W}}\left(\mathbf{w}\right) & = \lvert\widetilde{\pmb{\Sigma}}\rvert^{-\frac{1}{2}}\cdot M_{N}\cdot \int_{-\infty}^{\mathbf{w}_{(1)}}\cdot\cdot\cdot \int_{-\infty}^{\mathbf{w}_{(N)}} f_{\mathbf{DLN}}\left(\lvert\lvert\mathbf{w}-\widetilde{\pmb{\mu}}\rvert\rvert_{\widetilde{\pmb{\Sigma}}}\right)\  d\mathbf{\widehat{w}}_{(N)}\cdot\cdot\cdot d\mathbf{\widehat{w}}_{(1)} \\
\end{split}
\end{equation}
which characterizes a general elliptical multi-variate DLN RV.

Finally, note that the scaling matrix $\widetilde{\pmb{\Sigma}}$ is not the covariance matrix of $\mathbf{W}$ due to the heavy-tails of $\mathbf{W}$, similar to other heavy-tailed elliptical distributions such as the multi-variate Stable, t, or Laplace distributions. Further note that the normalization integral in Equation~\ref{eq:fR} is numerically unstable for high values of N (e.g., $N\geq 5$), and care should be taken when deriving the PDF of high-dimensional DLN RVs.

\section{Methods for heavy-tailed analysis}
\label{sec:Methods}

As discussed above, a main difficulty of working with the DLN distribution stems from its ``double exponential'' nature, i.e. the fact it exhibits exponential tails in both the positive and negative directions. The usual mitigation for a single exponential tail, applying a log transform, fails as the log is undefined on the negatives. This section describes how to extend methods applied to one-sided exponential tails to double-exponential distributions.

\subsection{Inverse-Hyperbolic-Sine space and the ADLN}

A common alternative to using log-transforms is transforming the data using the Inverse Hyperbolic Sine (asinh). For a review of the use of asinh in economic applications see \cite{BellemareWichman2020}. The hyperbolic sine and its inverse are given by
\begin{equation} \label{eq:ASINH}
\begin{split}
\text{sinh}(x) & = \frac{e^{x}-e^{-x}}{2} \\
\text{asinh}(x) & = \log\left(x+\sqrt{1+x^2}\right)
\end{split}
\end{equation}
The asinh transform has the following useful properties:
\begin{enumerate}
    \item Differentiable and strictly increasing in x.
    \item $\text{asinh}(x)\approx \text{sign}(x)(\log\lvert x\rvert + \log(2))$, with the approximation error rapidly vanishing as $\lvert x\rvert$ increases.\footnote{About 1\% approximation error at $\lvert x\rvert$=4, and about 0.1\% at $\lvert x\rvert$=10.}
    \item Odd function, such that $\text{asinh}(-x) = -\text{asinh}(x)$.
    \item Zero based, such that $\text{asinh}(0)=0$
\end{enumerate}
I.e., asinh is a bijection similar in flavor to the neglog transform:
\begin{equation} \label{eq:NEGLOG}
\text{neglog}(x) =\text{sign}(x)\log(1+\lvert x\rvert)
\end{equation}
but with less distortion than the neglog around 0, at the cost of the fixed bias $\log(2)\approx 0.7$.

It is useful to note that any difference of exponentials function can be factored into an exponential multiplied by a Hyperbolic Sine, i.e., 
\begin{equation} \label{eq:NEGLOG}
y =\exp\left(x_1\right) - \exp\left(x_2\right) = 2\cdot\exp\left(\frac{x_1 + x_2}{2}\right)\cdot\text{sinh}\left(\frac{x_1 - x_2}{2}\right)
\end{equation}
which highlights the intimate intuitive relation between the sinh function and the DLN and Laplace distributions. All three are expressed in terms of difference of exponentials, leading to their characteristic ``double exponential'' nature. Sinh's inverse, the asinh, is hence a natural transform to apply to DLN and Laplace distributed RVs.

As asinh is differentiable and strictly increasing, the method of transformation applies. If $Z=\text{asinh}(W)$ where $W\sim DLN$ then $Z\sim ADLN$, $W=\text{sinh}(Z)$, and $\frac{dZ}{dW} = \left(1+\text{sinh}(Z)^{2}\right)^{-1/2}$. We can now write the PDF for the ADLN distribution
\begin{equation} \label{eq:PDFADLN}
f_{ADLN}(z) = \frac{f_{DLN}(\text{sinh}(z))}{\text{asinh}'(\text{sinh}(z))} = f_{DLN}(\text{sinh}(z))\sqrt{1+\text{sinh}(z)^2} 
\end{equation}
which allows analysis of $Z\sim ADLN$, the transformed DLN RVs, whose histogram is more ``compact'' and easier to present.

Panels (c) and (d) of Figure~\ref{fig:DLNexam} present typical DLN distributions encountered in practice with linear (Panel c) and asinh (Panel d) horizontal axis. Panel (c) presents a truncated segment of the distribution. Due to the asinh transform, Panel (d) is able to present the entire distribution. The approximate log-Normality of the positive and negative sides of the DLN is not visible in Panel (c), but is made clear by the asinh transform in Panel (d).

\subsection{Growth in DLN-distributed variates}
\label{sec:Growth}

How does one measure growth in DLN-distributed RVs? A firm that had $\$100M$ of income in year $1$ and $\$120M$ of income in year $2$ has certainly grown its income. One can argue whether it is preferable to say the firm grew by $\frac{120M}{100M}-1=0.2=20\%$ or by $\log(120M)-\log(100M)=0.182$ log-points, yet the question itself is well-formed. But what if the firm had $-\$100M$ of income (i.e., loss) in year $1$, and then $\$120M$ of income in year $2$? What was its growth? This section aims to provide a rigorous answer to that question.

To begin, we require a definition of growth. \cite{BarroSala-I-Martin2003} and \cite{StudenyMeznik2013} define instantaneous growth of a time-continuous and \emph{strictly positive} RV $Z(t)>0$ as 
\begin{equation} \label{eq:pergrowth}
\frac{dZ(t)/dt}{Z(t)} = \frac{Z'(t)}{Z(t)} \approx \frac{Z_{t+1}-Z_t}{Z_t}
\end{equation}
with the second part of the equation using the first-difference of discrete variables as an approximation to the derivative $Z'(t)$, which yields the well-known formulation of percentage growth in discrete variables. Generalizing this definition to $Z(t)\in \mathbb{R}$ yields:
\begin{equation} \label{eq:pergrowth2}
d\% \equiv \frac{dZ(t)/dt}{\lvert Z(t)\rvert} = \frac{Z'(t)}{\lvert Z(t)\rvert} \approx \frac{Z_{t+1}-Z_t}{\lvert Z_t\rvert} \ \ \text{for} \ \ Z(t) \neq 0
\end{equation}
which guarantees that $Z_{t+1}>Z_t$ will imply positive growth, regardless of the sign of $Z_t$. The approximate term $\left(Z_{t+1}-Z_t\right)/\lvert Z_t\rvert$ is  \emph{generalized percentage growth} (hereafter denoted d\%), and is explosive if $\lvert Z_t\rvert\to 0$, similar to ``traditional'' percentage growth.

Next, it is instructive to consider the growth of a log-Normally distributed RV, as most measures of size encountered in firm dynamics (and elsewhere) are approximately log-Normally distributed. To that end, consider the following setting:
\begin{equation} \label{eq:AR_LN}
\begin{split}
& X_{t+1} = \left(1-\rho_X\right)\cdot\mu_X + \rho_X\cdot X_{t} + \epsilon^{X}_{t} \\
& \epsilon^{X}_{t} \sim \mathcal{N}(0,\sigma_{X}^2) \\
& Y_{t} = \text{exp}\left(X_{t}\right)
\end{split}
\end{equation}
In which $X_{t}$ is a simple $AR(1)$ stochastic process, and hence distributes Normally, and $Y_{t}>0$ is log-Normally distributed. What is the growth in $Y_{t}$?

Applying the definition, we have:
\begin{equation} \label{eq:loggrowth}
\frac{Y'(t)}{\lvert Y(t)\rvert} = \frac{Y(t)\cdot X'(t)}{Y(t)} = X'(t) \approx X_{t+1} - X_t = \log(Y_{t+1}) - \log(Y_{t}) \equiv \text{dlog}(Y_{t+1})
\end{equation}
which yields the well-known formulation of growth as a difference in logs between consecutive values, denoted dlog(). The difference between Equations~\ref{eq:pergrowth2} and~\ref{eq:loggrowth} is in whether we differentiate before applying the first-difference approximation. Note that using percentage growth as in Equation~\ref{eq:pergrowth2} in this case would yield:
\begin{equation} \label{eq:pergrowthYt}
\frac{Y_{t+1}}{Y_t} - 1 = \exp(X_{t+1} - X_t) - 1
\end{equation}
or the general observation that percentage growth is a convex transform of log growth. It is further worth noting that $\lim_{\rho_X \to 1} \left(X_{t+1} - X_t\right) = \epsilon^{X}_{t}$. Log growth yields the innovation in the underlying AR(1) process, while percent growth yields the transformed value $\exp(\epsilon^{X}_{t})-1$. I.e., percent growth introduces a convexity bias relative to log growth in the case of a log-Normally distributed RV.

Conversely, using log growth to measure growth in a Normally distributed RV, even if said RV is strictly positive in practice, would introduce a similar but opposite ``concavity bias.'' To see that, consider the growth in $X(t)>0$, when measured in dlog terms:
\begin{equation} \label{eq:loggrowthNorm}
\text{dlog}(X_{t+1}) = \log(X_{t+1}) - \log(X_t) = \log\left(\frac{X_{t+1}}{X_t} -1 +1\right) =  \log\left(\frac{ X'(t)}{\lvert X(t)\rvert} + 1\right)
\end{equation}
Put differently, using dlog() to measure growth in $X$ yields the log of percent growth, which is the appropriate measure by the definition in Equations~\ref{eq:pergrowth} and~\ref{eq:pergrowth2}. Hence, the concept of growth used is closely related to the distribution being considered.

Next, consider a similar setting, but for a DLN RV:
\begin{equation} \label{eq:AR_DLN}
\begin{split}
& X^{p}_{t+1} = \left(1-\rho_{p}\right)\cdot\mu_{p} + \rho_{p}\cdot X_{t}^{p} + \epsilon^{p}_{t} \\
& X^{n}_{t+1} = \left(1-\rho_{n}\right)\cdot\mu_{n} + \rho_{n}\cdot X_{t}^{n} + \epsilon^{n}_{t} \\
& (\epsilon^{p}_{t},\epsilon^{n}_{t})^{T} \sim \mathcal{N}\left(\pmb{0},\pmb{\Sigma}\right) \\
& Y^{p}_{t} = \text{exp}\left(X^{p}_{t}\right) \ \ ; \ \ Y^{n}_{t} = \text{exp}\left(X^{n}_{t}\right) \\
& W_{t} = Y^{p}_{t} - Y^{n}_{t}
\end{split}
\end{equation}
with $\pmb{\Sigma}$ as in Equation~\ref{eq:BVN}. By applying the generalized growth definition~\ref{eq:pergrowth2}, we have:
\begin{equation} \label{eq:DLNGROWTH}
\begin{split}
\frac{W'(t)}{\lvert W(t)\rvert} & = \frac{Y^{p}(t)\cdot dX^{p}(t)/dt - Y^{n}(t)\cdot dX^{n}(t)/dt}{\lvert W(t)\rvert} \approx \frac{Y^{p}_{t}\cdot\left(X^{p}_{t+1} - X^{p}_{t}\right) - Y^{n}_{t}\cdot\left(X^{n}_{t+1} - X^{n}_{t}\right)}{\lvert W(t)\rvert} \\
& = \frac{Y^{p}_{t}\cdot\text{dlog}\left(Y^{p}_{t+1}\right) - Y^{n}_{t}\cdot\text{dlog}\left(Y^{n}_{t+1}\right)}{\lvert Y^{p}_{t} - Y^{n}_{t}\rvert}
\end{split}
\end{equation}
which implies the growth of a DLN RV can be defined as a function of the levels and growth rates of its two component log-Normal RVs. Section~\ref{sec:MC} conducts Monte-Carlo experiments to explore the relation between the measures of growth presented above for Normal, log-Normal, and DLN distributed RVs.

\comments{
\begin{equation} \label{eq:DLNGROWTH}
\begin{split}
g & = \frac{200\cdot\left(\log\left(270\right)-\log\left(200\right)\right) - 100\cdot\left(\log\left(120\right)-\log\left(100\right)\right)}{\lvert200 - 100\rvert} = 0.4179 \ dlnp\  (\approx 0.4055 \ lp)  \\
g & = \frac{50\cdot\left(\log\left(270\right)-\log\left(50\right)\right) - 100\cdot\left(\log\left(120\right)-\log\left(100\right)\right)}{\lvert50 - 100\rvert} = 1.3218 \ dlnp
\end{split}
\end{equation}
}

\section{Monte-Carlo experiments}
\label{sec:MC}

This section reports the results of Monte-Carlo experiments designed to ascertain the properties of the moments, estimators, and measures discussed above. \comments{, as well as present further results on the properties of the DLN as an approximating distribution.}

\subsection{Properties of estimators}

I begin by exploring the moments and parameter estimators of Sections~\ref{sec:Moms} and~\ref{sec:Estim}. I concentrate the experiments on a region of the parameter space that arises in practical applications related to the theory of the firm:
\begin{equation} \label{eq:MC_Region_1}
\pmb{Q}: \ \ \left(\mu_p,\sigma_p,\mu_n,\sigma_n,\rho_{pn}\right) \in \left(\left[-3,3\right],\left[0.5,2.5\right],\left[-3,3\right],\left[0.5,2.5\right],\left[-1,1\right]\right)
\end{equation}

\noindent The data collection/creation for the Monte-Carlo analysis proceeds as follows.\\
\noindent For each $i \in \{1...N\}$:
\begin{enumerate}
    \item Draw a parameter vector $\pmb{\Theta}_i\in\pmb{Q}$ with Uniform probability.
    \item Calculate the theoretical central moments based on $\pmb{\Theta}_i$ using the method of Section~\ref{sec:Moms}.
    \item Draw $K$ observations $W_{i,k}\sim\text{DLN}(\pmb{\Theta}_i)$.
    \item Calculate the first five empirical central moments of $W_{i,k}$.
    \item Recalculate the first five empirical moments using iteratively smaller subsets of the $K$ observations.\footnote{Specifically, I recalculate the moments based on the first $K/2^s$ observations for $s\in\{1...11\}$.}
    \item Estimate the parameters of $W_{i,k}$, denoted $\pmb{\widehat{\Theta}}_i$, using the method of Section~\ref{sec:Estim}.
    \item Calculate the Kolmogorov-Smirnov (K-S), Chi-square (C-2), and Anderson-Darling (A-D) test statistics based on $\pmb{\widehat{\Theta}}_i$ and $W_{i,k}$.
\end{enumerate}
I repeat the data creation process $N=70,000$ times. Within each loop, I draw $K=100,000$ observations $W_{i,k}\sim\text{DLN}(\pmb{\Theta}_i)$.

Panel (a) of Table~\ref{tab:MC1} presents the Monte-Carlo results for the moment estimators of Section~\ref{sec:Moms}. It compares the theoretical moments derived in Step 2 of the Monte-Carlo experiment to the empirical moments derived in Step 4, concentrating on the first five moments of the distribution. The analysis is done in asinh space because the moments of the DLN explode quickly due to its heavy tails (similar to moments of the log-Normal, which are similarly considered in log space). The empirical and theoretical moments show high correlation, and the odd moments (mean or $1^{st}$ moment, skewness or $3^{rd}$ moment, and $5^{th}$ moment) exhibit no significant bias. The even moments (variance or $2^{nd}$, and kurtosis or $4^{th}$) show evidence of bias, which is fairly severe for kurtosis. Small-sample bias correction to the kurtosis estimator appears warranted, but is outside the scope of this work. The IQR of the difference between the theoretical and empirical moments is increasing with the moment degree, as expected.

% Estimator Monte-Carlo
\RPprep{Estimator Monte Carlo Experiments}{0}{0}{MC1}{%
    This table presents results of estimator Monte-Carlo experiments with $N=70,000$ repetitions and $K=100,000$ observations drawn in each repetition. Panel (a) tests the moments estimators $\widehat{M}_i\ \ i\in\{1...5\}$ of Section~\ref{sec:Moms} vs. the actual moments $M_i$, conducting all analysis in asinh space. It reports the general accuracy corr($\text{asinh}(\widehat{M}_i),\text{asinh}(M_i)$); the bias median($\text{asinh}(\widehat{M}_i)-\text{asinh}(M_i)$) ; and the accuracy IQR($\text{asinh}(\widehat{M}_i)-\text{asinh}(M_i)$). Panel (b) reports similar statistics comparing the DLN parameter estimators of Section~\ref{sec:Estim} $\pmb{\widehat{\Theta}}$ and the actual parameters $\pmb{\Theta}$. Panel (c) reports the values of parameters a,b,c,d in the approximations $ICDF(p) = a\cdot\exp(b\cdot p) + c\cdot\exp(d\cdot p)$ for the ICDFs of the Kolmogorov-Smirnov, Chi-square, and Anderson-Darling test statistics for DLN RVs, as well as the approximation $R^2$.
}
\RPtab{%
    \begin{tabularx}{\linewidth}{Frrrrr}
    \toprule
	\textit{Panel (a): Moment estimators} & $\widehat{M}_1$ & $\widehat{M}_2$ & $\widehat{M}_3$ & $\widehat{M}_4$ & $\widehat{M}_5$ \\
    \midrule
    Correlation                 &  0.9997  &  0.9929 &  0.9282 &  0.8238 &  0.8478 \\
    Bias                        &  -0.0001 &  0.1092 & -0.0002 &  6.3410 & 0.0220 \\
    Accuracy                    &  0.0217  &  0.4785 &  3.4480 &  8.5609 & 32.0236 \\ \\
    
	\textit{Panel (b): Parameter estimators} & $\widehat{\mu}_p$ & $\widehat{\sigma}_p$ & $\widehat{\mu}_n$ & $\widehat{\sigma}_n$ & $\widehat{\rho}_{pn}$ \\
    \midrule
    Correlation                 &  0.9408 &  0.9619 &  0.9412 &  0.9623 &  0.9190 \\
    Bias                        & -0.0034 &  0.0019 & -0.0043 &  0.0019 & -0.0048 \\
    Accuracy                    &  0.0588 &  0.0251 &  0.0614 &  0.0259 &  0.0762 \\ \\

    \textit{Panel (c): ICDF approximations} & a & b & c & d & $R^2$ \\
    \midrule
    Kolmogorov-Smirnov          & 6.75e-7 &  0.1553 & -6.7520 & -0.0011 &  0.9976 \\
    Chi-square                  & 1.88e-8 &  0.1955 &  1.2080 &  0.0044 &  0.9920 \\
    Anderson-Darling            & 1.18e-5 &  0.1350 & -5.7070 & -0.0060 &  0.9900 \\
	\bottomrule
    \end{tabularx}
}

Panel (b) of Table~\ref{tab:MC1} goes on to present the Monte-Carlo results for the parameter estimators of Section~\ref{sec:Estim}. It compares the actual parameters drawn in Step 1 to the estimated parameters calculated in Step 6. The results indicate the estimation procedure is performing quite well. There is high correlation between the actual and estimated parameters, including the hard to estimate correlation parameter. The parameter estimates also exhibit no systematic bias and reasonably low estimation error IQR. These results imply the estimation procedure, while cumbersome, is able to capture the DLN parameters correctly.

To further explore the precision and small-sample bias of the moment estimators, Figure~\ref{fig:MC1} presents the dependence of estimator quality on sample size. Panel (a) of the figure presents the dependence of the correlation between the theoretical and empirical moments on sample size. Kurtosis is even less precise than the $5^{th}$ moment, and is strongly influenced by sample size. Panel (b) of Figure~\ref{fig:MC1} then presents the dependence of the bias on sample size. The $1^{st}$ and $3^{rd}$ moment estimators exhibit no small-sample bias. The $2^{nd}$ and $5^{th}$ exhibit small and rapidly decreasing bias. Kurtosis, again, shows high bias, only slowly decreasing with sample size.

% Estimator Monte Carlo 
\RPprep{Estimator Monte-Carlo experiments}{0}{1}{MC1}{%
    This figure presents results of estimator Monte-Carlo experiments. Panel (a) graphs the dependence of the correlation between the theoretical and empirical moments on sample size. Panel (b) graphs the dependence of moment bias on sample size. Panel (c) presents the distribution of (log of) the K-S statistic in the simulations. Panels (d)-(f) then present the ICDF of the (log) K-S, C-2, and A-D statistics, along with the fitted curves.
}
\RPfig{%
	\begin{tabular}{ccc} 
		\subfigure[Corr($\text{asinh}(\widehat{M}_i),\text{asinh}(M_i)$)] {\includegraphics[width=2.5in]{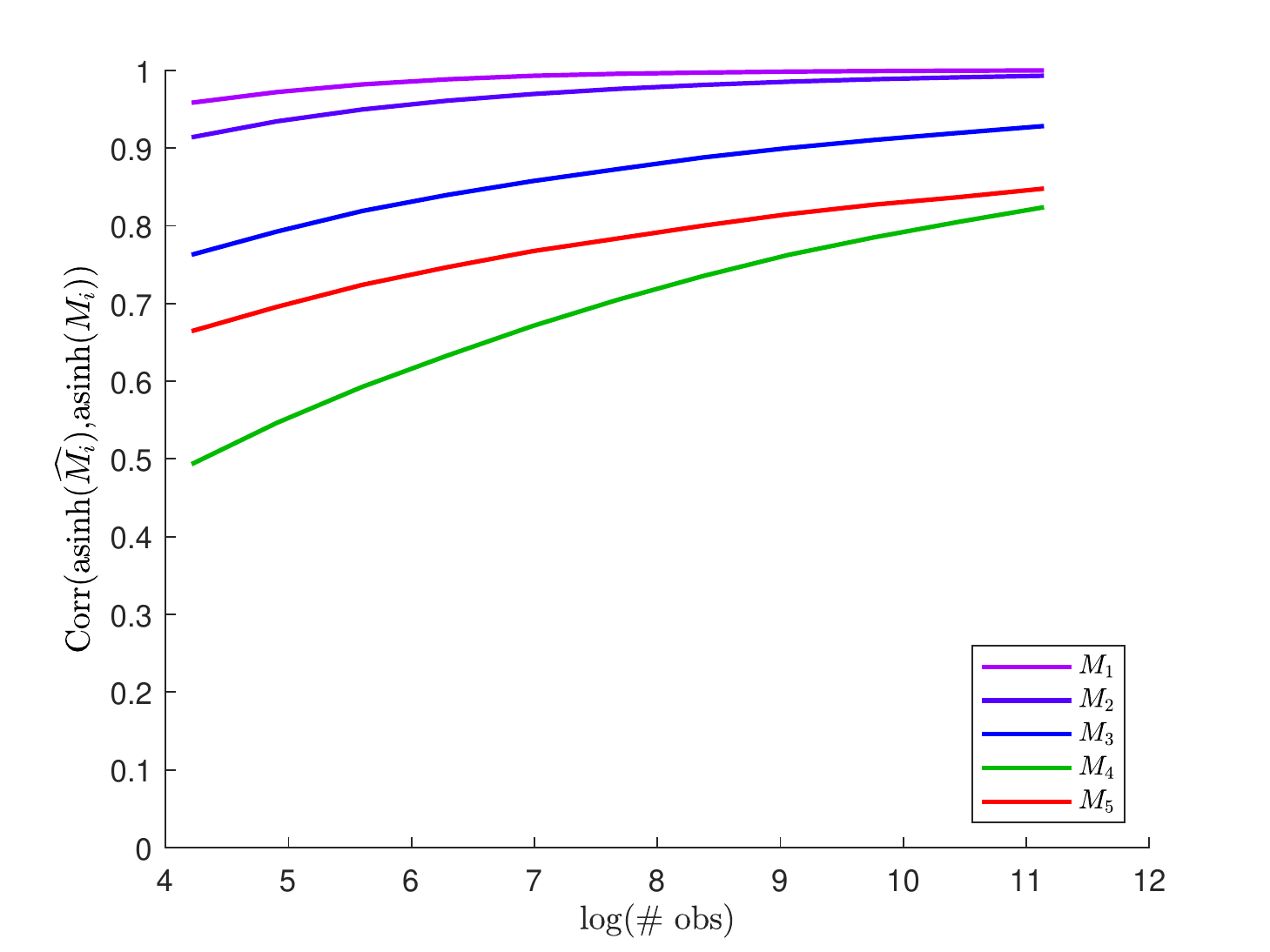}} & 
		\subfigure[Median($\text{asinh}(\widehat{M}_i)-\text{asinh}(M_i)$)] {\includegraphics[width=2.5in]{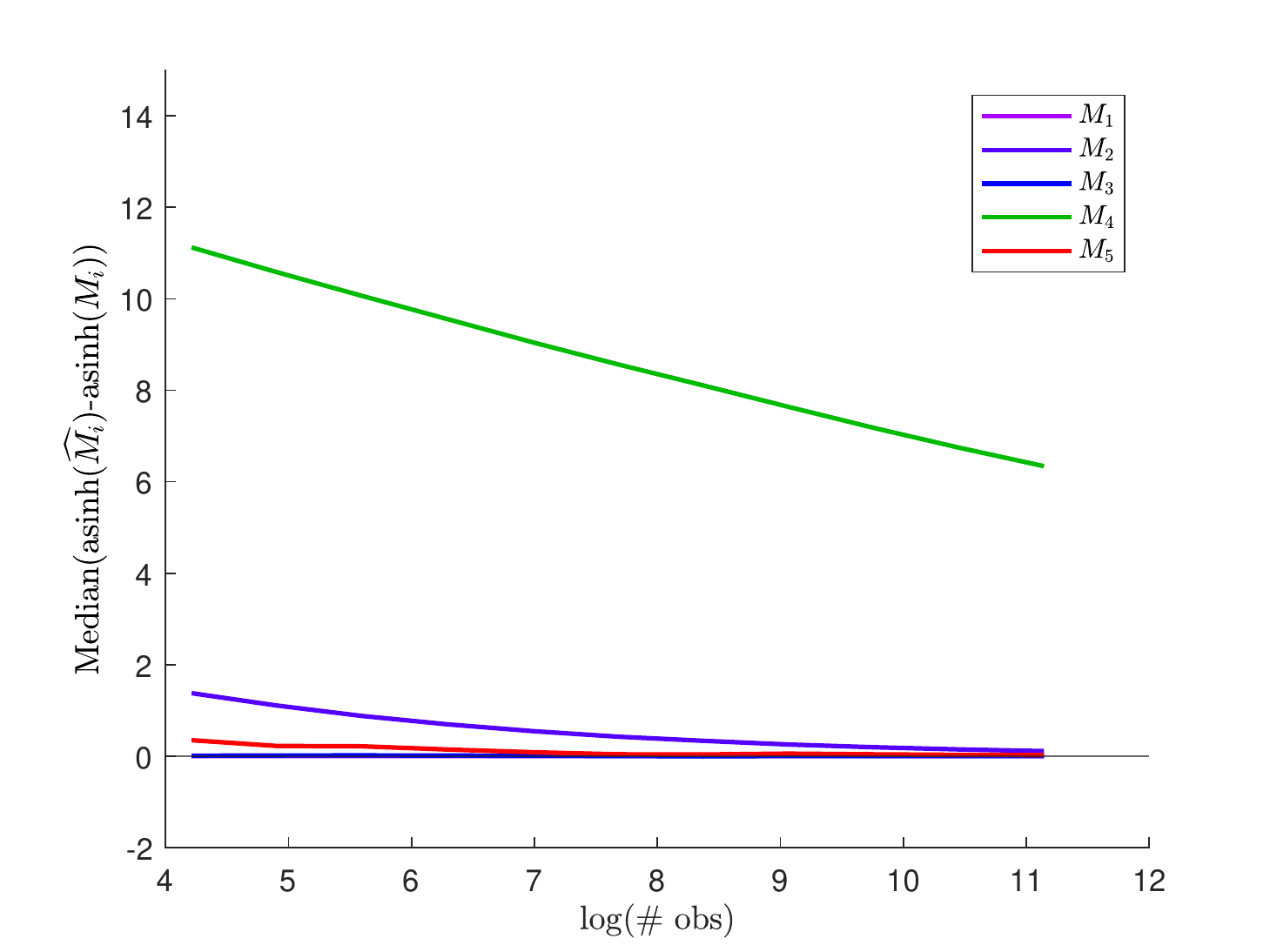}} & 
		\subfigure[PDF of log K-S statistic] {\includegraphics[width=2.5in]{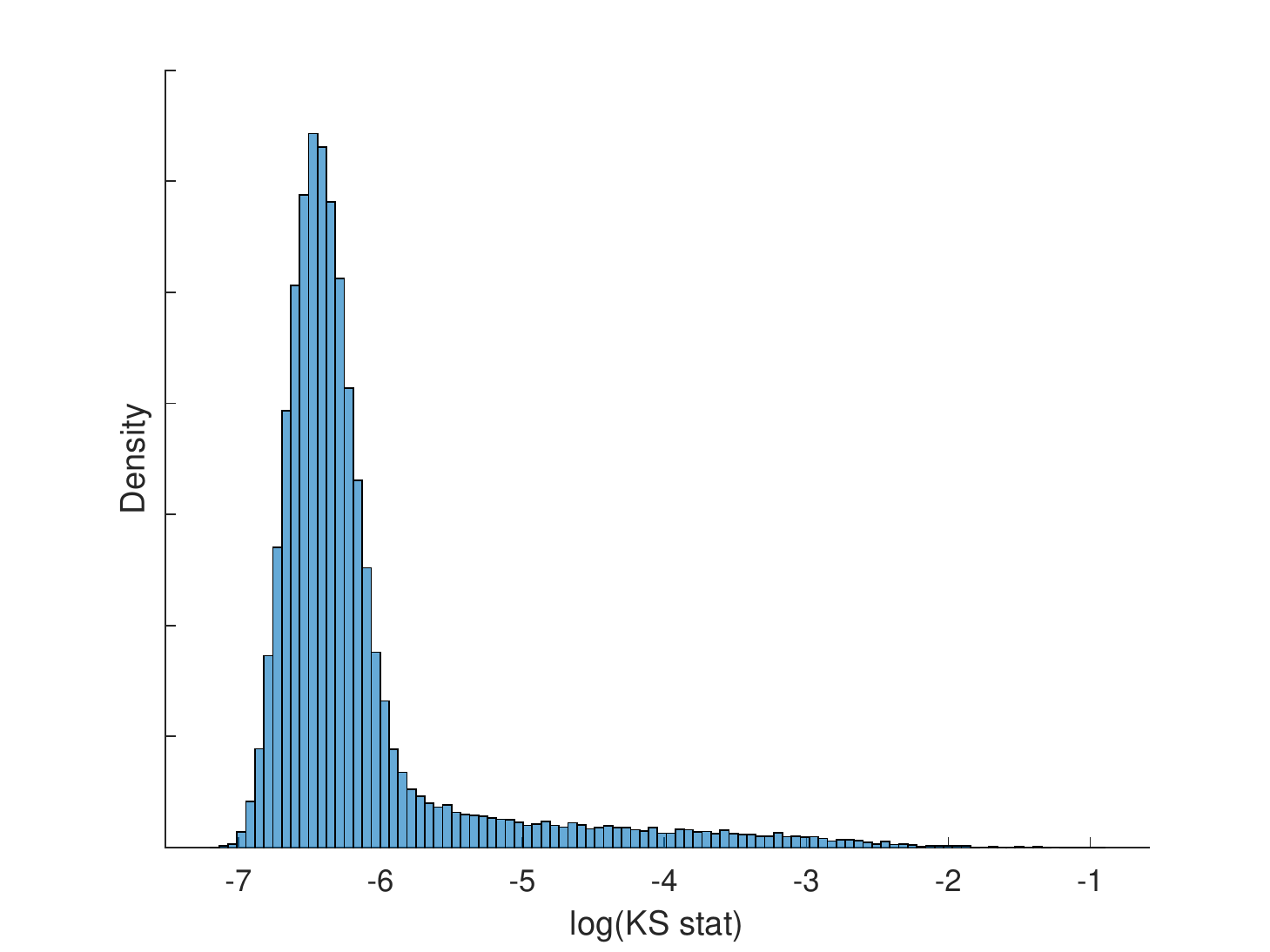}} \\ \\
		\subfigure[ICDF of log K-S statistic]
		{\includegraphics[width=2.5in]{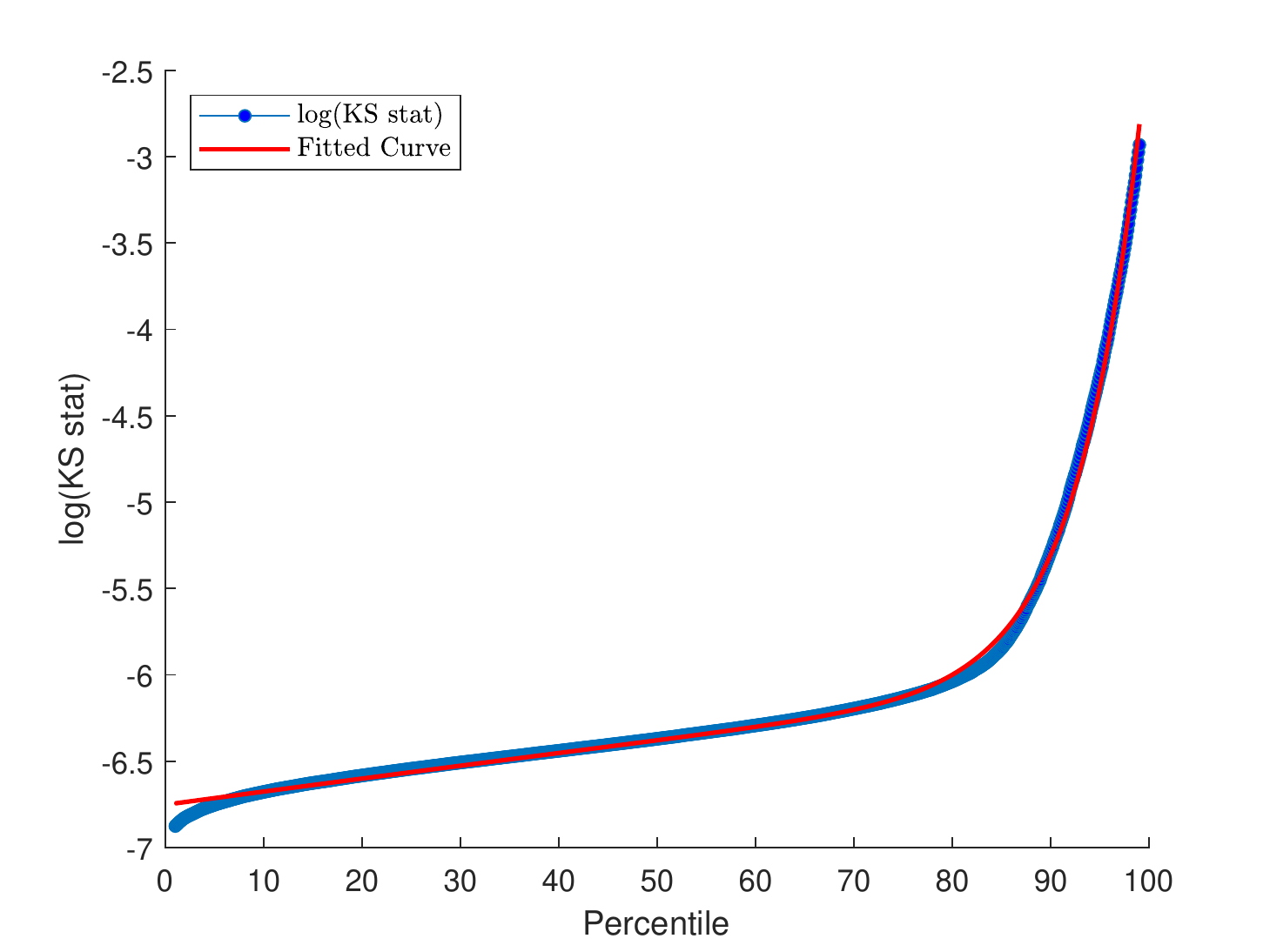}} &
		\subfigure[ICDF of log C-2 statistic]
		{\includegraphics[width=2.5in]{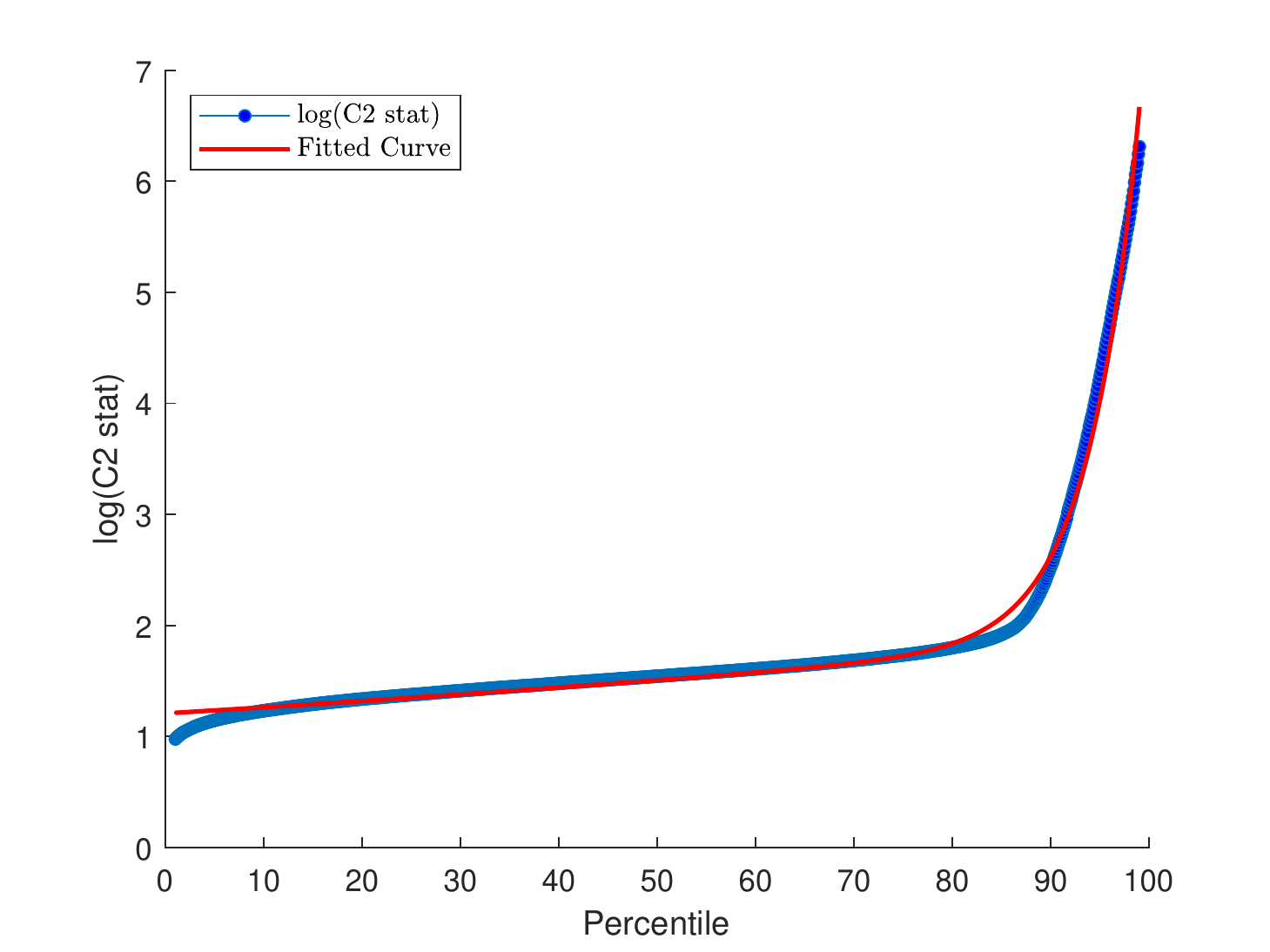}} &
		\subfigure[ICDF of log A-D statistic]
		{\includegraphics[width=2.5in]{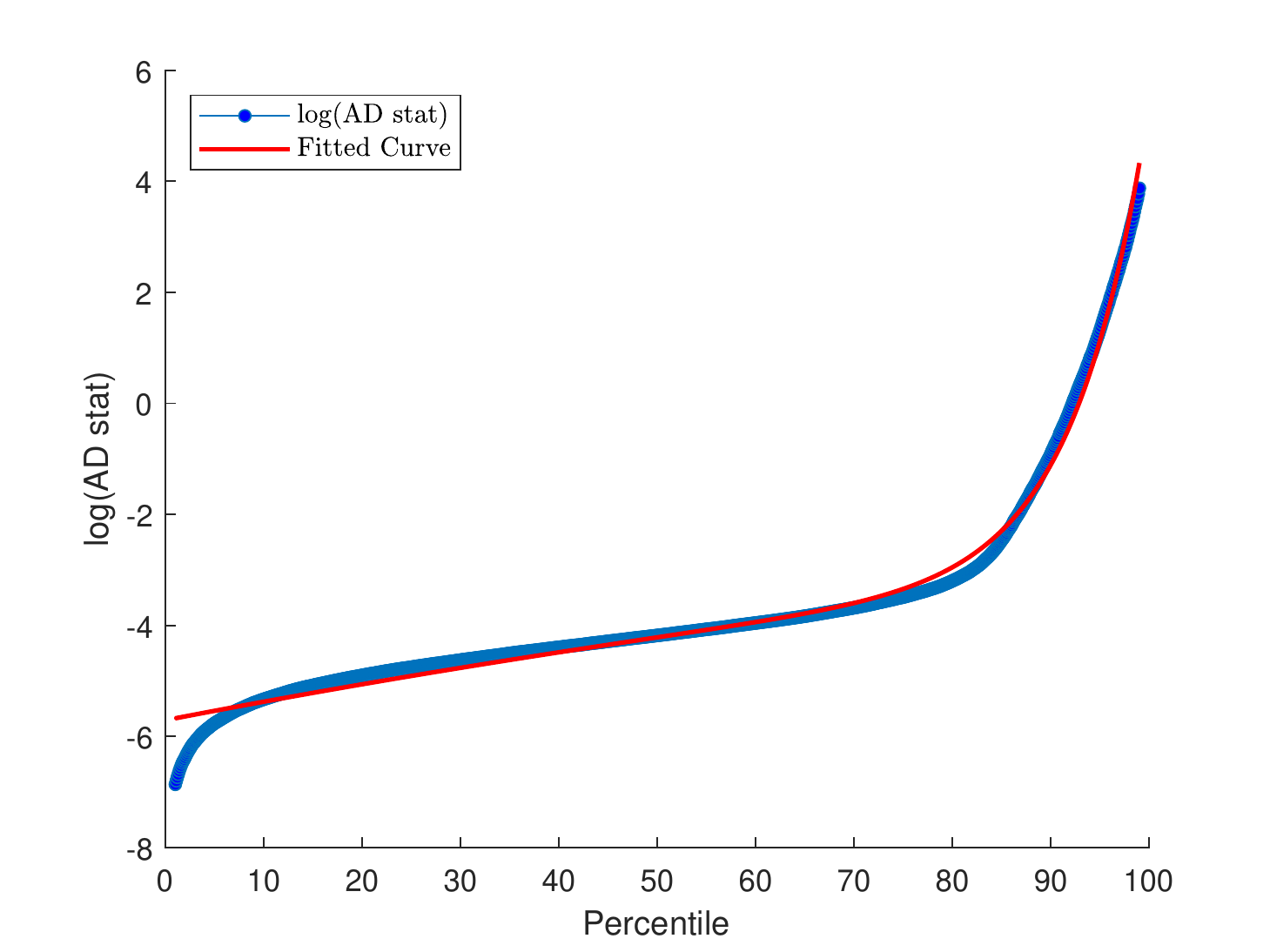}} \\ \\
	\end{tabular}
}

\subsection{Test-statistic critical values}
\label{sec:TestStats}

A second goal of the Monte-Carlo experiments is to establish critical values for test statistics of the hypothesis that some given data are drawn from a DLN distribution. This is especially important for the Anderson-Darling test statistic, whose critical values are well-known to strongly depend on the distribution being examined. See e.g. \cite{Stephens1979}, \cite{DAgostinoStephens1986} Chapter 4, and \cite{JantschiBolboaca2018}.

To that end, I calculate the K-S, C-2, and A-D test statistics for each of the $N$ draws in the sample, as described in Step 8 above. To fix ideas, Panel (c) of Figure~\ref{fig:MC1} presents the distribution of (log of) the K-S statistic in the Monte-Carlo experiment. I then calculate the inverse-CDF (ICDF) of the resulting distribution of (log of) each test statistic. Panels (d), (e), and (f) of Figure~\ref{fig:MC1} present the ICDFs of the (log) K-S, C-2, and A-D test statistics, respectively. E.g., Panel (f) indicates that one should reject the hypothesis that given data are drawn from the DLN distribution (at a 5\% confidence level) if the A-D statistic is higher than $\text{exp}(ICDF(95)) = \text{exp}(1.135) = 3.110$.

To move from calculating critical values to deriving a continuous mapping between p-values and test-statistic values, it is common in the literature discussed above to propose an ad-hoc functional form which is able to approximate the ICDF well. Once one estimates the approximating functional form using non-linear least-squares, one can use it to find the p-values associated with each test-statistic value, and vice-versa. Following experimentation, the functional form most closely able to replicate the resulting ICDFs is of the form:
\begin{equation} \label{eq:Pvals}
ICDF(p) = a\cdot\text{exp}\left(b\cdot p\right) + c\cdot\text{exp}\left(d\cdot p\right)
\end{equation}
which is a four-parameter sum (or difference, if $c<0$) of exponentials.

Panels (d), (e), and (f) of Figure~\ref{fig:MC1} include the fitted values of the functional form, and show that there is an excellent fit between the functional form and the empirical ICDFs. Panel (c) of Table~\ref{tab:MC1} presents the values of the four approximating parameters for each of the (log) test statistics' ICDFs, and further reports the $R^2$ of the fit, which is above $0.99$ for all three statistics. Hence, one can safely use these functionals to derive p-values for tests of distributional hypotheses.

%In unreported results, I repeat the MC process and the estimation of approximating functional forms for the test statistic ICDFs of the Normal, skew-Normal, Stable, and Laplace distributions. This is done to facilitate tests comparing given data from unknown distributions to those distributions, in an equal manner to tests vs. the DLN. Thus, putting all distributions on equal footing and assuaging concerns that the DLN is preferentially treated in such tests. The resulting p-value functionals are included in the accompanying code suite.

\subsection{Growth measures}

A second set of Monte-Carlo experiments tests the relation between the growth measures described in Section~\ref{sec:Growth}, for RVs distributed Normal, log-Normal, and DLN. To that end, I define three stochastic processes yielding stationary distributions distributed N, LN, and DLN. For each RV type, in each Monte-Carlo iteration, I draw random parameters for the distribution, simulate it forward, measure growth per-period using the different measures discussed above, and consider the relation between the random innovations $\epsilon_t$ and the various growth measures.

The stochastic processes for $X$, $Y$, and $W$, distributed N, LN, and DLN, respectively, are as described in Equations~\ref{eq:AR_LN} and~\ref{eq:AR_DLN} above. The parameter regions are:
\begin{equation} \label{eq:MC_Region_2}
\begin{split}
\pmb{Q}_{N}: & \ \ \left(\rho_{N},\mu_{N},sd_{N}\right) \in \left(\left[0.60,0.99\right],\left[-100,100\right],\left[10,100\right]\right) \\
\pmb{Q}_{LN}: & \ \ \left(\rho_{LN},\mu_{LN},sd_{LN}\right) \in \left(\left[0.60,0.99\right],\left[-3,3\right],\left[0.5,2.5\right]\right) \\
\pmb{Q}_{DLN}: & \ \ \left(\rho^{p,n}_{DLN},\mu^{p,n}_{DLN},sd^{p,n}_{LN},\rho^{pn}_{DLN}\right) \in \left(\left[0.60,0.99\right],\left[-3,3\right],\left[0.5,2.5\right],\left[-1,1\right]\right) \\
\end{split}
\end{equation}
with $\sigma_{\Box} = \sqrt{sd_{\Box}^2\cdot\left(1-\rho_{\Box}^2\right)}$ for $\Box \in \{N,LN,DLN\}$.

\noindent The data collection/creation for the second Monte-Carlo analysis proceeds as follows:\\
\noindent For each RV type $\Box \in \{N,LN,DLN\}$: \\
\noindent For each $i \in \{1...N\}$:
\begin{enumerate}
    \item Draw a parameter vector $\pmb{\Theta}_i\in\pmb{Q}_\Box$ with Uniform probability.
    \item Initialize the RV $Z_{\Box,0}$ to $\mu_\Box$ for N, exp($\mu_\Box$) for LN, and $Z^{p,n}_{\Box,0}$ at exp($\mu^{p,n}_{\Box,0}$) for DLN.
    \item Draw a shock vector of length $K+100$ (two correlated shock vectors for DLN).
    \item Simulate the process forward $K+100$ period based on its laws of motion.
    \item Drop the first 100 observation as burn-in.
    \item Calculate the set of growth measures from Section~\ref{sec:Growth}.
\end{enumerate}
I repeat the data creation process $N=10,000$ times, each for $K=1,000$ periods, yielding a total of $10M$ growth observations to be analyzed per distribution type.

Panels (a),(b),(c) of Table~\ref{tab:MC2} presents the correlations between different growth measures for N, LN, and DLN RVs, respectively. The panels also report correlations concentrating on strictly positive values (i.e., when $Z_{t}>0$ and $Z_{t+1}>0$) and when further avoiding tiny beginning values (i.e., $Z_t>1$). The appropriate concept of growth for Normally distributed RV is $\epsilon_t/\lvert Z_{t}\rvert$, and Panel (a) shows it is highly correlated with the generalized percentage growth measure. The panel further shows that using dlog as a measure of growth for Normal RVs is inaccurate. This fact is further highlighted by Panels (a) and (b) of Figure~\ref{fig:MC2} which present the relation between the appropriate growth measure and the generalized percent (d\%) and dlog measures, respectively. Panel (a) shows d\% captures growth of Normal RVs well, and Panel (b) highlights the ``concavity bias'' arising from using the dlog measure rather than the d\% measure. The dispersion around the 45-degree line in Panel (a) is driven by the mean-reversion term of the AR(1), which the growth concept ignores.

% Growth Monte-Carlo
\RPprep{Growth Monte Carlo Experiments}{0}{0}{MC2}{%
    This table presents results of growth Monte-Carlo experiments with $N=10,000$ repetitions and $K=1,000$ observations simulated forward in each repetition. Panels (a), (b), and (c) present results for N, LN, DLN, respectively. Within each panel, I report correlations between the following measures of growth: $\epsilon_t$ the stochastic innovation underlying the growth at time $t$; $\epsilon_t/\lvert Z_{t}\rvert$ the relative stochastic innovation; d\%($Z_{t+1}$)=$\left(Z_{t+1} - Z_{t}\right/\lvert Z_{t}\rvert$ the generalized percentage growth; dlog($Z_{t+1}$)=log($Z_{t+1}$)-log($Z_{t}$) the log point growth; dDLN($Z_{t+1}$) the DLN growth formulation based on Equation~\ref{eq:DLNGROWTH}.
}
\RPtab{%
    \begin{tabularx}{\linewidth}{Flllll}
    \toprule
	\textit{Panel (a): N} & $\epsilon_t$ & $\epsilon_t/\lvert Z_{t}\rvert$ & d\%($Z_{t+1}$) & dlog($Z_{t+1}$) & \\
    \midrule
    $\epsilon_t$ 
    & 1.000 & 0.010 & 0.009 & 0.659$^{a}$ & \\
    $\epsilon_t/\lvert Z_{t}\rvert$ 
    & 0.380$^{b}$ & 1.000 & 0.973 & 0.031$^{a}$ & \\
    d\%($Z_{t+1}$) 
    & 0.357$^{b}$ & 0.960$^{b}$ & 1.000 & 0.033$^{a}$ & \\
    dlog($Z_{t+1}$) 
    & 0.712$^{b}$ & 0.590$^{b}$ & 0.617$^{b}$ & 1.000 & \\ \\

	\textit{Panel (b): LN} & $\epsilon_t$ & $\epsilon_t/\lvert Z_{t}\rvert$ & d\%($Z_{t+1}$) & dlog($Z_{t+1}$) & \\
    \midrule
    $\epsilon_t$ 
    & 1.000 & 0.023$^{a}$ & 0.269$^{a}$ & 0.931$^{a}$ & \\
    $\epsilon_t/\lvert Z_{t}\rvert$
    & 0.644$^{b}$ & 1.000 & 0.097$^{a}$ & 0.023$^{a}$ & \\
    d\%($Z_{t+1}$) 
    & 0.381$^{b}$ & 0.363$^{b}$ & 1.000 & 0.295$^{a}$ & \\
    dlog($Z_{t+1}$) 
    & 0.929$^{b}$ & 0.620$^{b}$ & 0.381$^{b}$ & 1.000 & \\ \\

	\textit{Panel (c): DLN} & $\widehat{\epsilon}_t^c$ & $\widehat{\epsilon}_t/\lvert Z_{t}\rvert^c$ & d\%($Z_{t+1}$) & dlog($Z_{t+1}$) & dDLN($Z_{t+1}$) \\
    \midrule
    $\widehat{\epsilon}_t^c$ 
    & 1.000 & 0.000$^{a}$ & 0.000$^{a}$ & 0.038$^{a}$ & 0.000$^{a}$ \\
    $\widehat{\epsilon}_t/\lvert Z_{t}\rvert^c$
    & 0.043$^{b}$ & 1.000 & 0.652 & 0.022$^{a}$ & 0.944 \\
    d\%($Z_{t+1}$) 
    & 0.009$^{b}$ & 0.464$^{b}$ & 1.000 & 0.016$^{a}$ & 0.645 \\
    dlog($Z_{t+1}$) 
    & 0.057$^{b}$ & 0.739$^{b}$ & 0.397$^{b}$ & 1.000 & 0.023$^{a}$ \\
    dDLN($Z_{t+1}$)
    & 0.040$^{b}$ & 0.931$^{b}$ & 0.455$^{b}$ & 0.797$^{b}$ & 1.000 \\ \\
    \bottomrule
    \end{tabularx}
    \begin{flushleft}
    $^a$ For strictly positive values ($Z_{t}>0$ and $Z_{t+1}>0$) \\
	$^b$ For strictly positive and non-tiny initial values ($Z_{t}>1$ and $Z_{t+1}>0$) \\
	$^c$ For DLN, I define $\widehat{\epsilon}_t = \left(Z_t^p\cdot\epsilon_t^p - Z_t^n\cdot\epsilon_t^n\right)$ and $Z_t = Z_t^p - Z_t^n$
    \end{flushleft}
}

% Growth Monte-Carlo
\RPprep{Growth Monte-Carlo experiments}{1}{0}{MC2}{%
    This figure presents results of growth Monte-Carlo experiments. Panels (a) and (b) graph the relation between the growth of a Normal RV and (a) generalized percentage growth d\%($Z_{t+1}$)=$\left(Z_{t+1} - Z_{t}\right)/\lvert Z_{t}\rvert$; (b) log point growth dlog($Z_{t+1}$)=log($Z_{t+1}$)-log($Z_{t}$). Panels (c) and (d) graph the relation between the growth of a log-Normal RV and (c) dlog($Z_{t+1}$) ; (d) d\%($Z_{t+1}$). Panels (e) and (f) graph the relation between the growth of a DLN RV and (e) the DLN growth measure from Equation~\ref{eq:DLNGROWTH}, dDLN($Z_{t+1}$); (f) d\%($Z_{t+1}$).
}
\RPfig{%
	\begin{tabular}{cc} 
		\subfigure[N growth vs. d\%$^a$] {\includegraphics[width=2.5in]{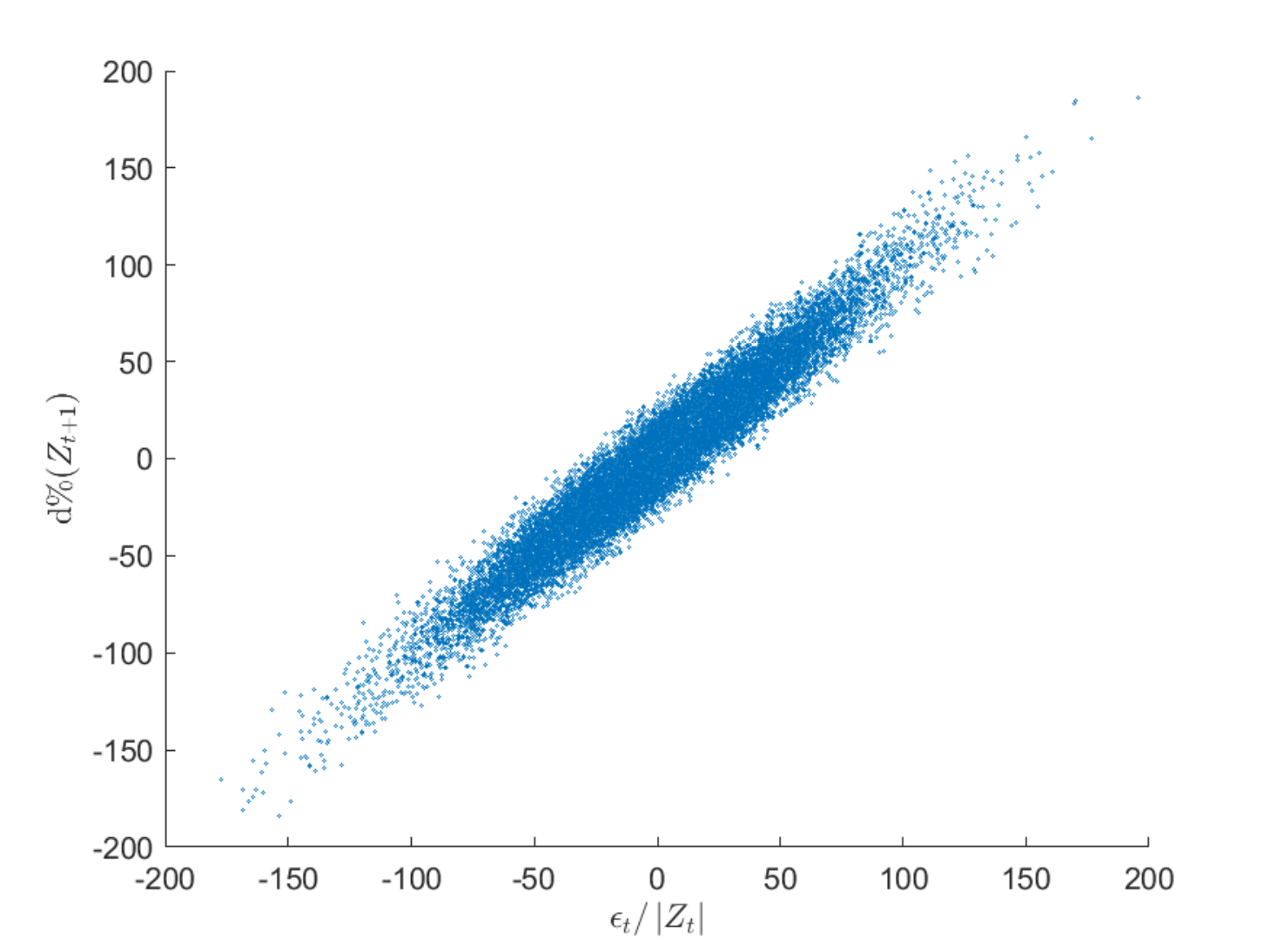}} & 
		\subfigure[N growth vs. dlog$^b$] {\includegraphics[width=2.5in]{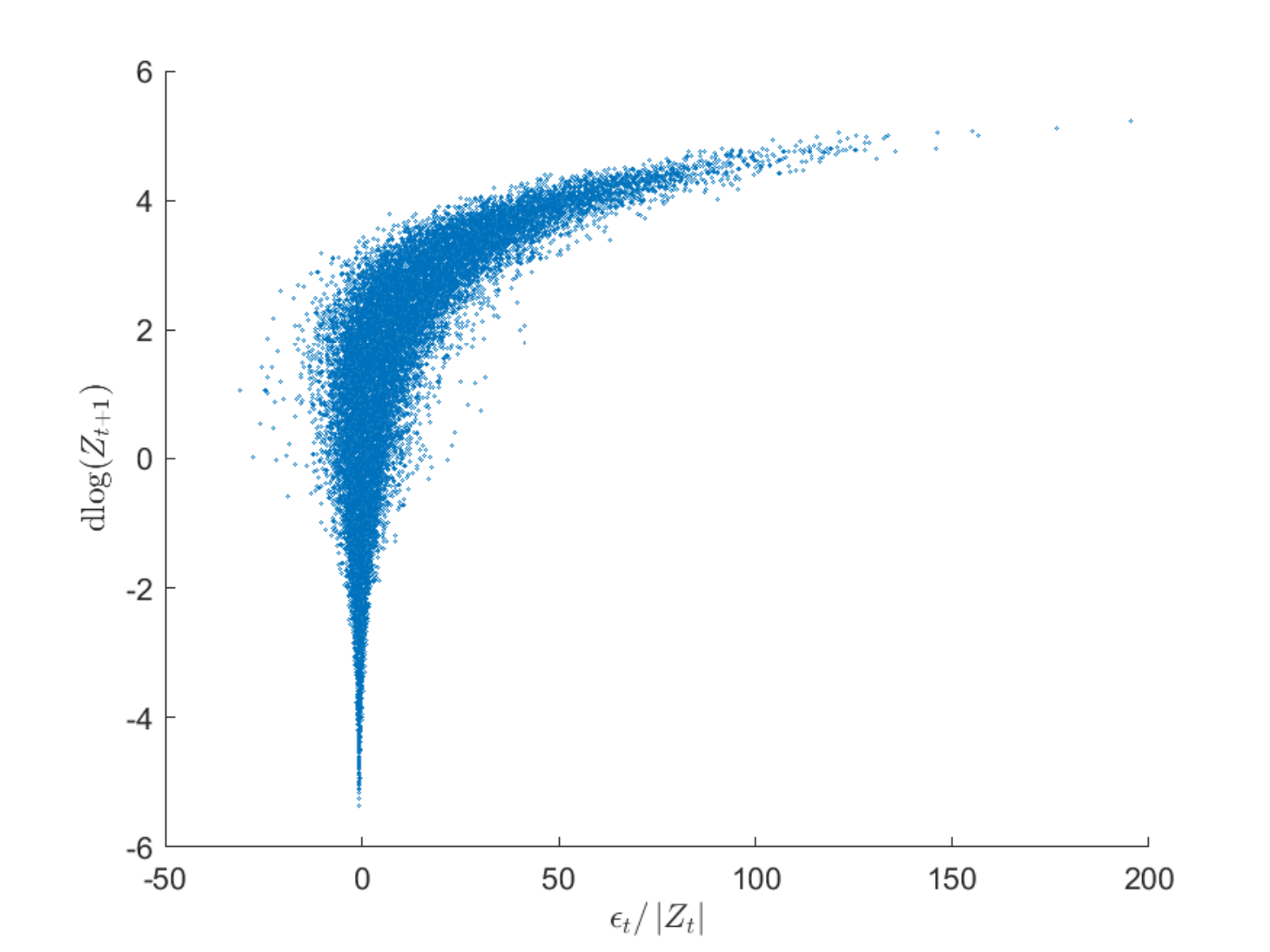}} \\ 
		\subfigure[LN growth vs. dlog] {\includegraphics[width=2.5in]{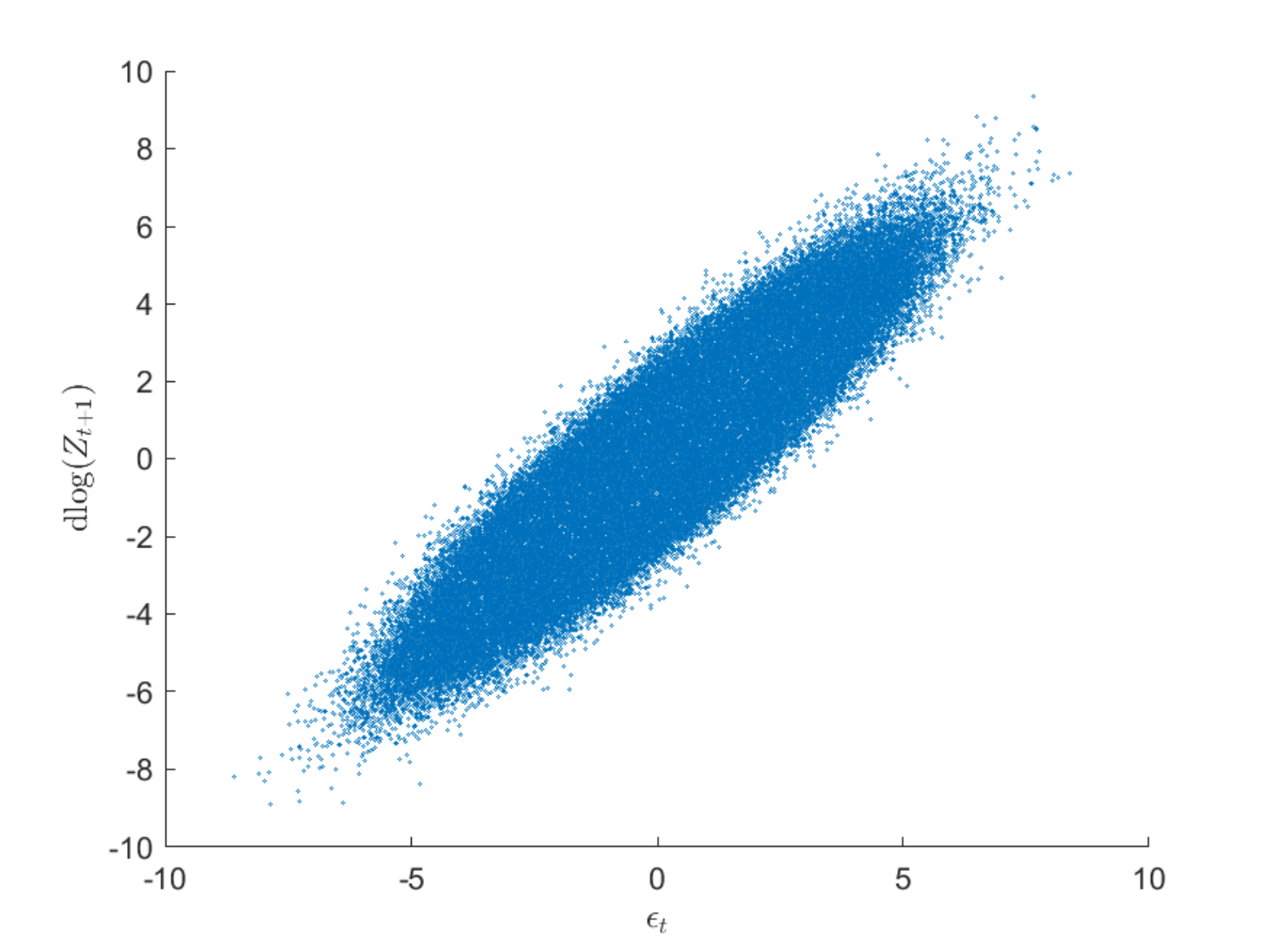}} & 
		\subfigure[LN growth vs. d\%] {\includegraphics[width=2.5in]{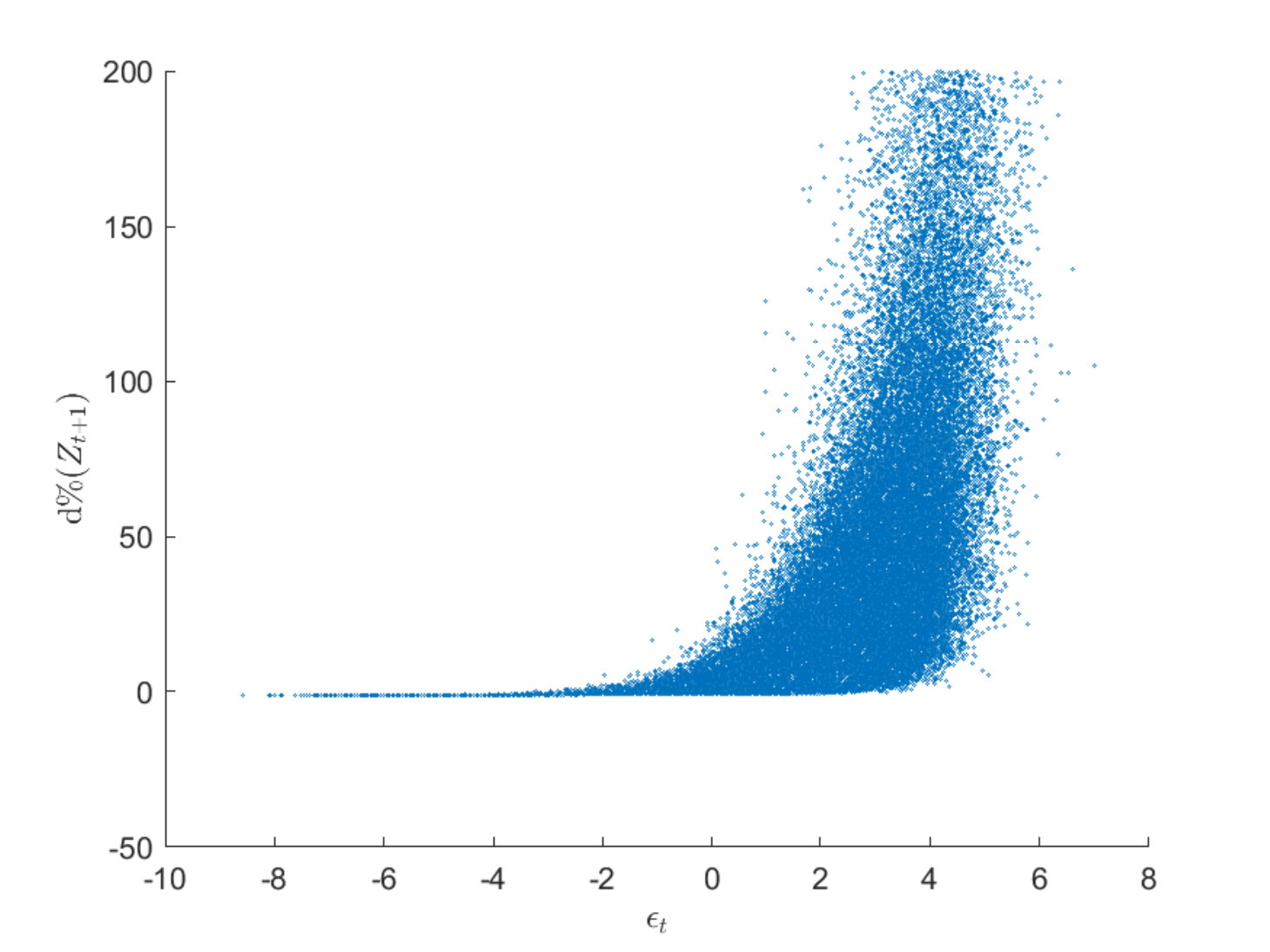}} \\ 		
		\subfigure[DLN growth vs. dDLN$^a$] {\includegraphics[width=2.5in]{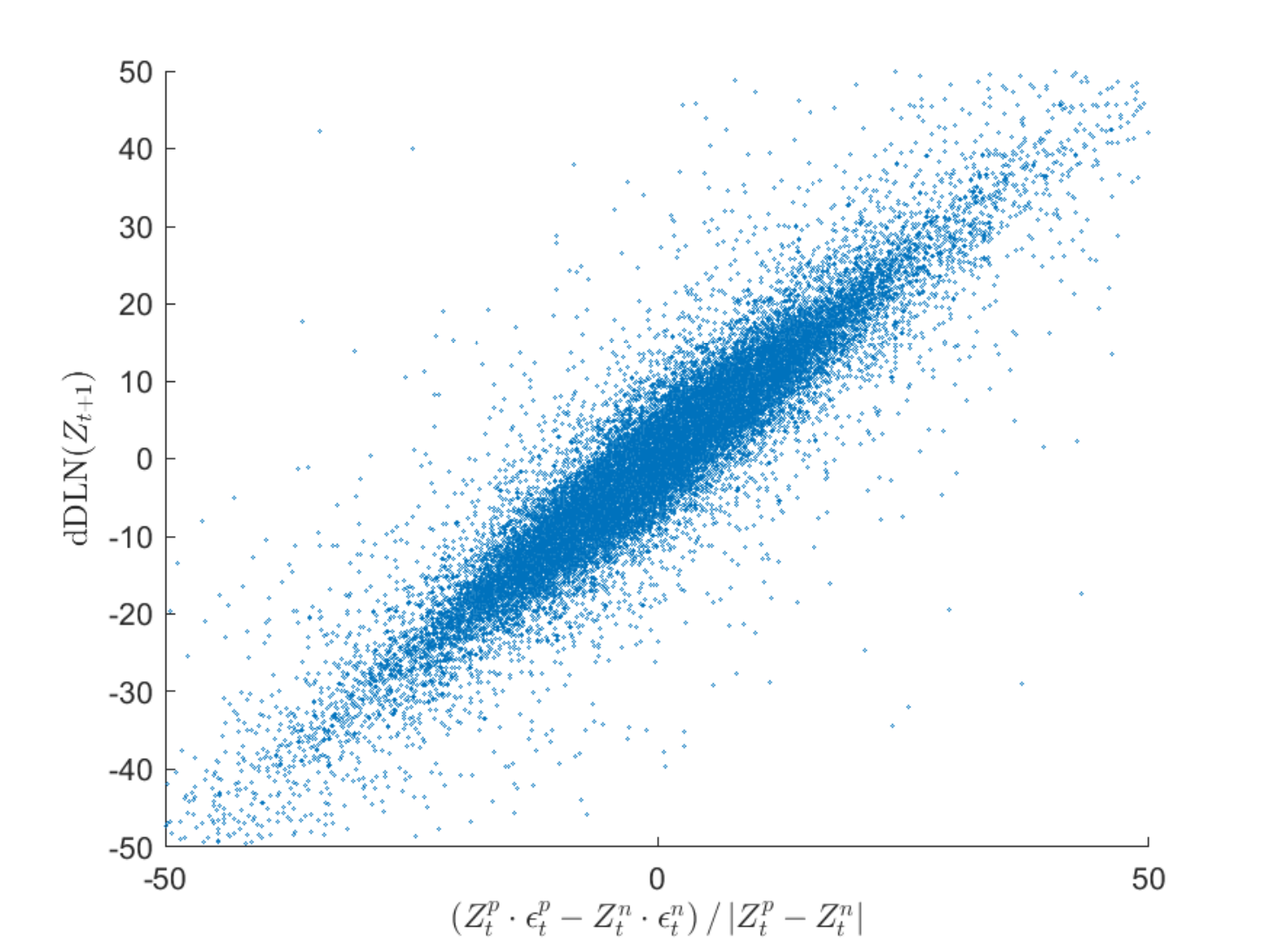}} & 
		\subfigure[DLN growth vs. d\%$^a$] {\includegraphics[width=2.5in]{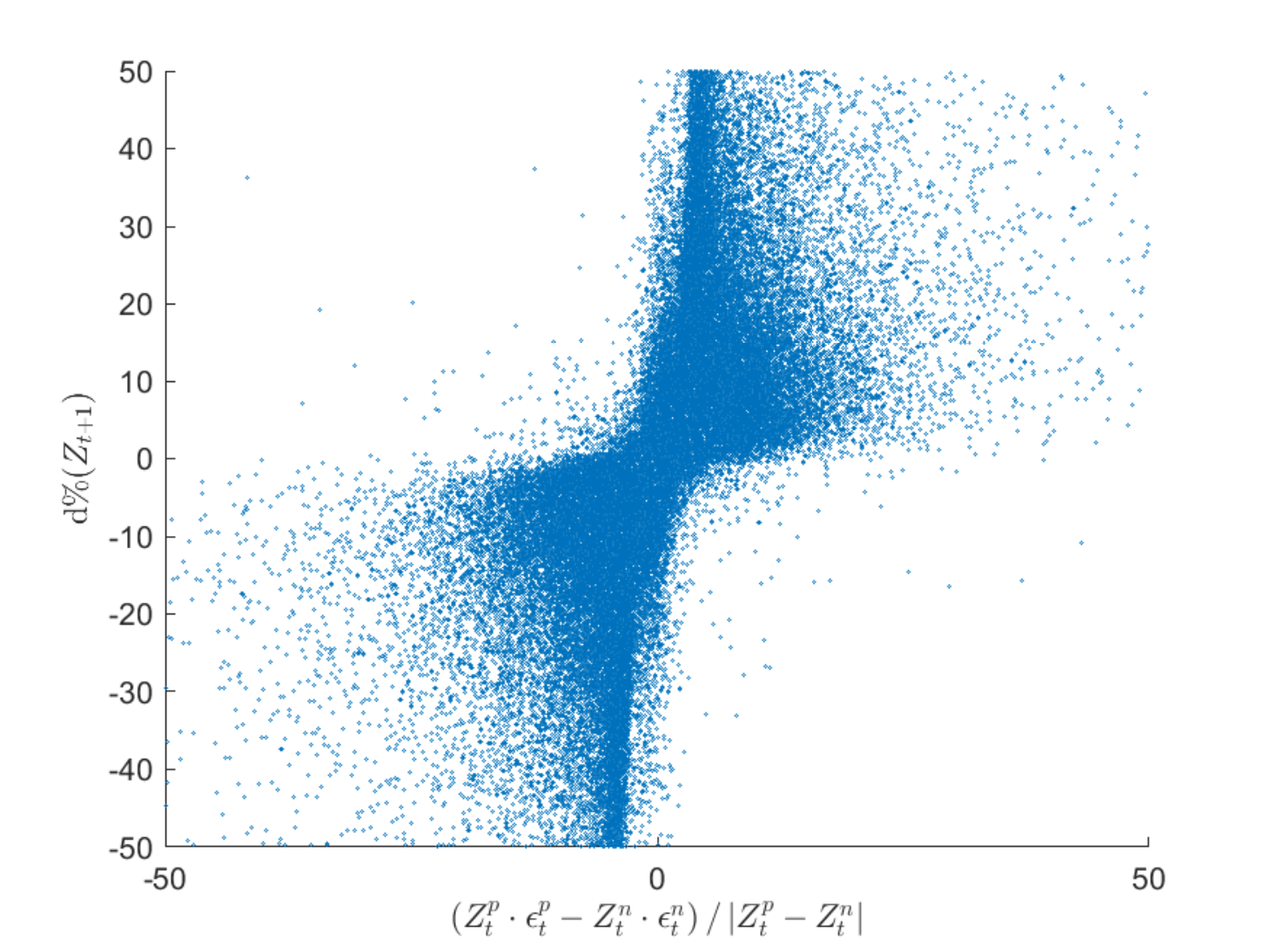}} \\ 
	\end{tabular}
    \begin{flushleft}
    $^a$ For non-tiny initial values ($\lvert Z_{t}\rvert>1$) \\
	$^b$ For strictly positive and non-tiny initial values ($Z_{t}>1$ and $Z_{t+1}>0$) \\
    \end{flushleft}	
}

Panel (b) of Table~\ref{tab:MC2} moves on to considering LN RVs. Here, the appropriate concept of growth is just $\epsilon_t$, and the panel shows that dlog measures growth well, while d\% suffers from a convexity bias and is a poor measure of growth. Panels (c) and (d) of Figure~\ref{fig:MC2} make the convexity bias clear by plotting the relation between growth and dlog and between growth and d\%, respectively.

Finally, Panel (c) of Table~\ref{tab:MC2} presents correlations between growth of DLN RVs and the growth measures. For DLN, the appropriate concept of growth is $\left(Z_t^p\cdot\epsilon_t^p - Z_t^n\cdot\epsilon_t^n\right)/\lvert Z_t^p-Z_t^n\rvert$, and the panel shows that the growth formula for DLN derived in Equation~\ref{eq:DLNGROWTH} captures it well. The panel also shows that dlog, which has limited usability for measuring DLN growth as it is limited to positive values, does poorly even when limited to positive values, and reaches a correlation of ~0.75 with DLN growth even when limiting to positive, non-tiny values. Panels (e) and (f) of Figure~\ref{fig:MC2} show that dDLN is indeed an appropriate measures, while d\% is an unbiased but noisy measure of DLN growth.

\section{Summary}

This paper presents the Difference-of-Log-Normals (DLN) distribution, stemming from the multiplicative CLT, and lays a methodological and quantitative foundation for the analysis of DLN-distributed phenomena. It begins by characterizing the distribution, defining its PDF and CDF, presenting estimators for its moments and parameters, and generalizing it to a elliptical multi-variate RVs.

It goes on to discuss mathematical methods useful in the analysis of DLN distributions. First, it shows the intimate intuitive relation between the DLN distribution and the Hyperbolic Sine, and why the Inverse Hyperbolic Sine (asinh) is a useful transform when dealing with ``double exponential'' RVs such as the DLN.

Next, it considers the concept of growth for DLN RVs. It extends the classical definition of growth, applying only to positive RVs, to RVs $\in \mathbb{R}$. It then shows that the measure of growth used is dependant on the distribution of the data being measured. It makes the case that growth in Normal, log-Normal and DLN RVs should be measured using different measures of growth and develops the appropriate measure of growth for DLN RVs.

The paper reports the results of extensive Monte-Carlo experiments, aimed to establish the properties of the estimators and measures presented. It shows that the moment estimators have good accuracy, but highlights their small-sample bias, especially for the case of kurtosis. A small-sample bias-correction method for the kurtosis estimator is merited. It also shows that the parameter estimators proposed are reasonably accurate and unbiased. To enable accurate tests of whether some data are DLN, it establishes critical values and p-value estimators for three distributional tests: Kolmogorov-Smirnov, Chi-square, and Anderson-Darling.

A second Monte-Carlo experiment verifies the generalized growth measures discussed indeed back-out the appropriate growth concept for Normal, log-Normal, and DLN distributions. It especially highlights the ``convexity/concavity bias'' arising when applying the wrong measure of growth to an RV. Of importance here is the evidence that measuring growth of log-Normal RVs using percentage growth leads to a significant convexity bias.

\clearpage
\bibliographystyle{JFE}
\bibliography{MyLibrary}

%\input{FigTab.tex}
%%%%%%%%%%%%%%%%%%%%%%%%%%%%%%%%%%%%%%%%%%%%%%%%%%%%%%%%%%%%%%%%%%%%%%%%%%%%%%%%%%%%%%%%%%%%%
% Appendices
\appendix

\renewcommand\thefigure{\thesection.\arabic{figure}}
\renewcommand\thetable{\thesection.\arabic{table}}
\renewcommand\theequation{\thesection.\arabic{equation}}
\setcounter{figure}{0}
\setcounter{table}{0}
\setcounter{equation}{0}

\clearpage

\end{document}